\documentclass[referee]{aa}
\usepackage{graphicx}
\usepackage[]{natbib}
\usepackage[]{lscape}
\bibpunct{(}{)}{;}{a}{}{,}
\newcommand{\etal}{{\it et al.}}

\newcommand{\kms}{\mbox{km~s$^{-1}$}}

\def\sec{\ifmmode{^{\prime\prime}}\else{$^{\prime\prime}$}\fi}
\def\min{\ifmmode{^{\prime}}\else{$^{\prime}$}\fi}
\def\deg{\ifmmode{^\circ}\else{$^\circ$}\fi}
\def\arcsec#1.#2 {\ifmmode {#1^{\prime\prime}\hskip-0.42em.
                  \hskip0.15em#2}
         \else {$#1^{\prime\prime}\hskip-0.42em.\hskip0.15em#2$}
         \fi}
\def\arcmin#1.#2 {\ifmmode {#1^{\hskip 0.05em\prime}\hskip-0.35em.
                  \hskip0.05em#2}
         \else {$#1^{\hskip 0.05em\prime}\hskip-0.35em.\hskip0.05em#2$}
         \fi}
\def\arcdeg#1.#2 {\ifmmode {#1\deg\hskip-0.42em.
                  \hskip0.10em#2}
         \else {$#1\deg\hskip-0.42em.\hskip0.10em#2$}
         \fi}

\begin{document}

\title{Properties of Isolated Disk Galaxies}

\author{J. Varela \inst{1}
           \and M. Moles \inst{1}
          \and I. M\'arquez\inst{1}
           \and G. Galletta\inst{2}
           \and J. Masegosa\inst{1}
           \and D. Bettoni\inst{3}
                         }

\offprints{M. Moles}

\institute{Instituto de Astrof\'{\i}sica de Andaluc\'{\i}a (C.S.I.C.)
               Apartado 3004, 18080 Granada, Spain\\
              \email{jesusv@imaff.cfmac.csic.es, moles@iaa.es, isabel@iaa.es, pepa@iaa.es}
              \and Dipartimento di Astronomia, Universit\`a di Padova, Vicolo
               dell'Osservatorio 2, I-35122, Padova\\
             \email{galletta@pd.astro.it}
         \and Osservatorio Astronomico di Padova, Vicolo dell'Osservatorio 5,
        I-35122, Padova\\
             \email{bettoni@pd.astro.it}
             }

\date{Received ; accepted  }

\titlerunning{Isolated Galaxies}

\abstract{A new sample of northern isolated galaxies, which are defined by the physical criterion that they were not affected by other galaxies in their evolution during the last few Gyr, is presented. To find them we used the logarithmic ratio, $f$, between inner and tidal forces acting upon the candidate galaxy by a possible {\it perturber}. The analysis of the distribution of the $f$-values for the galaxies in the Coma cluster lead us to adopt the criterion $f \leq -4.5$ for isolated galaxies. The candidates were chosen from the CfA catalog of galaxies within the volume defined by cz $\leq$5000 km/s, galactic latitude higher than 40\deg~ and declination $\ge -$2.5\deg. The selection of the sample, based on redshift values (when available), magnitudes and sizes of the candidate galaxies and possible 
perturbers present in the same field is discussed. The final list of selected 
isolated galaxies comprises of 203 objects from the initial 1706. The list contains 
only truly isolated galaxies in the sense defined, but it is by no means complete, 
since all the galaxies with possible companions under the $f$-criterion but with 
unknown redshift were discarded. We also selected a sample of perturbed galaxies 
comprising of all the disk galaxies from the initial list with companions (with 
known redshift) satisfying  $f \geq -$2 and $\Delta(cz) \le 500$ km/s, a total of 
130 objects. The statistical comparison of both samples shows significant differences
 in morphology, sizes, masses, luminosities and color indices. Confirming previous 
results, we found that late spiral, Sc type galaxies are, in particular, more 
frequent among isolated galaxies, whereas Lenticular galaxies are more abundant 
among perturbed galaxies. Isolated systems appear to be smaller, less luminous and 
bluer than interacting objects. We also found that bars are twice as frequent among 
perturbed galaxies compared to isolated galaxies, in particular for early Spirals 
and Lenticulars. The perturbed galaxies have higher L$_{FIR}$/L$_B$ and 
M$_{mol}$/L$_B$ ratios, but the atomic gas content is similar for the two samples. 
The analysis of the luminosity-size and mass-luminosity relations shows similar trends
 for both families, the main difference being the almost total absence of big, bright 
and massive galaxies among the family of isolated systems, together with the almost 
total absence of small, faint and low mass galaxies among the perturbed systems. All 
those aspects indicate that the evolution induced by interactions with  neighbors, 
would proceed from late, small, faint and low mass Spirals to earlier, bigger, more 
luminous and more massive spiral and lenticular galaxies, producing at the same time 
a larger fraction of barred galaxies, but preserving the same relations between global
 parameters. The properties we found for our sample of isolated galaxies appear rather 
similar to those of high redshift galaxies, suggesting that the present day isolated 
galaxies could be quietly evolved, unused {\sl building blocks} surviving in low 
density environments. 

}

\maketitle

\section{Introduction}

Galaxies are presently most commonly found in aggregates of different density. Inside these aggregates, the observed morphology and the segregation of types \citep{dressler, fas00} may have been determined by the local density of matter and by the subsequent interactions between galaxies. Exchanges and interaction with its neighbors can affect the galaxy global properties and their star formation rates, as shown by the members of dense groups or pairs \citep{moles}, even if the effects, in general, are not as dramatic as they are in the strongest cases (Moles et al., 1994b; M\'arquez \& Moles, 1999).

Another way to look into the effects of interaction on the overall equilibrium of a galaxy is through the analysis of the relations found between the structural parameters for different families of galaxies. The E/S0 cluster (Faber et al., 1987; Davis \& Djorgovski, 1987; J\o rgensen, Franx, \& Kj\ae rgaard, 1996), and field galaxies (Treu et al., 2001) satisfy the Fundamental Plane (FP) relation, even if its form is not the same for both families. On their side, the Spiral (S) galaxies satisfy the Tully-Fisher (TF) relation. It has been indicated that the TF relation defined by isolated spiral galaxies presents a smaller scatter than non-isolated galaxies (M\'arquez \& Moles, 1996; M\'arquez et al., 2002; van Kannappan et al., 2002).

To find isolated galaxies, a physically grounded and operational definition of isolation has to be given. Often one speaks of field galaxies to refer to systems that are not in dense aggregates, but frequently they are still under the influence of neighboring galaxies and cannot be considered as genuinely isolated objects (see for example M\'arquez et al., 2002). The first systematic compilation of isolated galaxies was made by Karachentseva (1973), who used as the main criterion that they don't show companions within 20 diameters in the PSS plates. Later, \cite{tg76}, while searching for groups of galaxies, found that some of the galaxies couldn't be assigned to any of the identified groups. It was argued that they would trace a cosmological homogeneously distributed population of galaxies. This conclusion was however challenged by \cite{ht77} who used a deeper sample (m $\leq$ 15.7 instead of 
m $\leq$ 14). The next step was taken by \cite{vsc86}, who adopted stricter criteria to find isolated galaxies. Their conclusion was that it was very unlikely that there were a cosmological component of galaxies without any kind of clustering (at least at small redshift). But an important observation was the detection of galaxies in such poor environments that they had very probably evolved without any external perturbation, as if they were isolated. In that perspective what is needed is an operational definition of isolation that could lead to identifying galaxies that have evolved free from external influences for most of its life, a point of view adopted by \cite{mm96} and M\'arquez et al. (1999). In this context, it was noticed that non-axisymmetric structures like bars, tails, or plumes, which are usually explained invoking gravitational interaction with companions or satellites, can also be present in isolated galaxies, posing the problem of their origin in absence of sizable interactions (Moles et al., 1994; Moles, M\'arquez \& P\'erez, 1995).

In the present work we discuss the criteria used to find isolated galaxies and build a statistically well defined sample to analyze their properties. Due to the lack of redshift information for most of the faint, possible {\it perturbers}, the final sample is by no means complete, our main goal being to keep only truly isolated galaxies, discarding all the suspect systems. In the last section we 
compare the properties of the family of the isolated galaxies with that of confirmed perturbed galaxies, to show the main differences between both families.

We also consider the similarity of some of the properties of present day isolated systems with those reported for the presumed {\sl building blocks} in the distant Universe \citep{ferg03,trujillo}. The point is that stellar systems situated in regions of very low density would have evolved differently from galaxies in aggregates without a significant contribution from the surroundings and, therefore they might be similar to those original building blocks.

\section{Isolation criteria}

From an operational point of view, we will consider a galaxy to be isolated when its evolution in terms of structure and stellar content has been dominated by internal forces for most of its life. Or, in  other terms, when the external forces are judged to be unable to have produced ponderable changes in it in at least the last 2$\times$10$^9$ years. As found by \cite{Atha84}, and \cite{byrd92}, an external interaction can only have influence on the structure of a given system when the corresponding tidal force amounts to $\geq 1\%$ of the internal force. 

The results of numerical simulations of encounters between galaxies can be described in terms of a tidal perturbation parameter (see Byrd \& Howard 1992). It gives the ratio between the tidal force exerted by the perturber, P, on the primary galaxy, G, and the internal force per unit mass in the outer parts of the primary. Given a galaxy of mass M$_G$ and size R, and a perturber of mass M$_P$ 
with a closest approach, pericenter distance given by b, that ratio is 
\begin{equation}
\frac{F_{ext}}{F_{int}}\propto \left(\frac{M_P}{M_G}\right)\times 
\left(\frac{R}{b}\right)^{3}
\label{nonso}
\end{equation}

In this expression the masses of the galaxies can be evaluated using the diameters or the luminosities. Actually, since there is a relation between luminosity and size, both parameters should lead to similar results. We preferred to use the magnitudes here instead of the sizes (used for instance by Dahari, 1984) since they are known for a larger number of galaxies and easier to correct. Indeed, it is assumed that M/L is reasonably similar for all the galaxies. Admittedly this is a rough hypothesis but appropriate since the final criteria for isolation we 
are going to adopt has enough room to be insensitive to the differences in M/L 
from galaxy to galaxy.

Regarding the pericenter parameter, b, Icke (1985), when studying the influence 
over the gas in the disk of a galaxy due to the flyby of another one, found that 
the maximum pericenter distance still able to trigger star formation in the gas 
disk, is given by
\begin{equation}
b = (4\pi g\frac{v}{s}\mu )^{1/3}r \label{terza}
\end{equation}
\noindent 
where {\it g} is a geometrical factor describing the encounter that ranges between 
0 to 1; {\it v} the gas speed at the distance {\it r} of the center of the disk; 
{\it s} the sound speed at the same distance; and $\mu$ the ratio of the perturber's 
mass over the central galaxy mass. The factor in brackets is, at most, of the order 
of a few, and therefore {\it b} is, at most, of the order of or a few times the 
radius of the perturbed galaxy. In other words, only close encounters could produce 
important effects in the internal dynamics of the gas in the other system, in 
agreement with the results by \cite{Atha84} and \cite{byrd92}.

Since the pericenter parameter b cannot be directly estimated, we adopted instead 
the projected distance D$_p$ between the galaxy and the perturber on the plane of 
the sky at the primary's distance. Thus, the final expression for the perturbation 
parameter is
\begin{equation}
f=\log\left(\frac{F_{ext}}{F_{int}}\right)=3\log\left(\frac{R}{D_p}\right) + 0.4\times 
(m_G - m_P)
\label{f}
\end{equation}
\noindent 
where m$_G$ and m$_P$ are the apparent magnitudes of the primary and perturber 
galaxies, respectively.

In fact $D_p$ can be greater than the impact parameter, $b$, depending on the orbit 
type, the orientation of the line of nodes with respect to the observer, and the 
position of the perturber on its orbit around the primary, all quantities totally 
unknown. In those cases a galaxy would be judged more isolated (i. e., smaller 
f-values) than it really is, which would compromise our selection criterion. The 
error in $f$ assuming $D_p$ instead of $b$ amounts to $3\times log(D_p/b)$. As we 
will see, given that we are going to make a statistical use of the f, and the 
restrictive limits we impose in order to consider a galaxy as isolated, the adopted 
criterion is robust enough and only in very extreme (i. e., very improbable) cases,
 a perturbed system might be considered as isolated.

The theoretical results by Athanassoula (1984), Byrd \& Howard (1992), and Icke (1985), consistently show that values of $f \ge -2$ are required to produce sizable effects on a given disk galaxy. However, before fixing the limit $f$-value to consider a galaxy isolated, it has to be taken into account the fact that we are using rather rough estimates for the masses of the galaxies and for the pericenter distance. Thus, systems with $f$-values observed now to be smaller than $-2$ could had have presented 
larger values in the past, depending on the details of the orbit and of the reaction 
to the interaction.

\begin{figure}
\resizebox{9cm}{!}{\includegraphics[angle=-90]{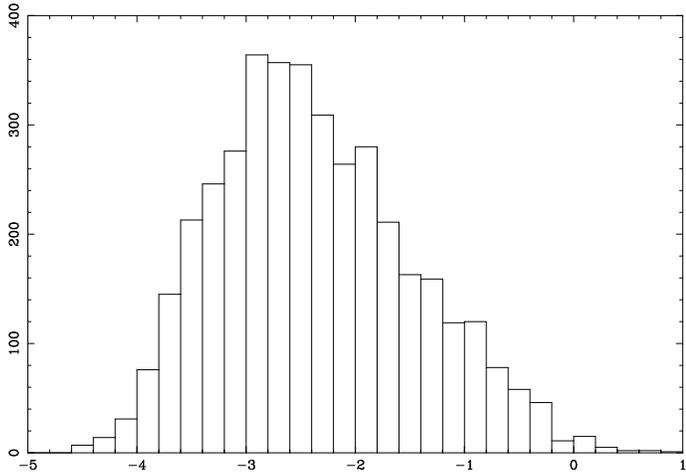}}
\caption{The distribution of the f-values for galaxies in the Coma Cluster}
\label{histcoma}
\end{figure}

To evaluate the typical $f$ values existing in well defined galaxy aggregates, we studied their distribution for the galaxies of the Coma Cluster. As far as it can be considered in a stationary state, the range of $f$-values we find at a given moment should be statistically representative of the values it can have along the time, and it would be possible to extract conclusions that are valid for the whole duration of the stationary state. Therefore we took data from \citet{coma83}, and considered all the galaxies in Coma with photometry in the B and R bands, a total of 4075 objects. They cover a wide range of magnitude and size. The distribution of the $f$-parameter values is presented in Fig.~\ref{histcoma}. The median value of the distribution is $-$2.7 and, as it can be seen in the Figure, there are no galaxies with $f < -$4.5.
%


In view of these results for Coma, it seems safe to adopt the formal criterion 
$f \leq -$4.5 to select what we'll call isolated galaxies.

From the preceding considerations it is clear that the $f$ cut-off value is probably too strict for situations where a candidate galaxy has only a few, at most, possible companions. Thus, in particular, it is robust enough to cope with probable errors in the different observable parameters entering the definition, and ensure that all the selected objects are actually isolated. From  expression (\ref{f}), it is simple to verify that errors of 20\% in size and/or luminosity translate into errors of only a few tenths in $f$ at most, still very far from critical values. Similarly, even a ratio between the projected distance and the pericenter distance as large as $D_P$/$b$ = 3, would produce an $f$-value greater than the {\it true} value by 1.4 units, still far from critical values. Only for very eccentric orbits, and when the perturber is around the apocenter, our criterion would fail and could select {\it false} isolated galaxies.

Another aspect to consider is the possibility of inducing biases in the family of 
the selected objects, given the dependency of $f$ on the luminosity and size. A 
galaxy might have a different probability of being taken as isolated, for a fixed 
population of possible companions. However, as far as there is a direct relation 
between size and luminosity, we expect only a small net effect on $f$. Thus, 
considering L$\propto$R$^2$, the difference in the $f$-values produced by a given
 perturber on galaxies with luminosities of the ratio 1/1.5  would amount to only 
0.17, well inside the uncertainties in M/L or any other parameter, and unable to 
approach the measured $f$ to critical values.

Given the preceding considerations we don't expect strong biases in the family of 
selected isolated galaxies.

\subsection{Comparison with previous criteria}

\cite{k73} considered a galaxy of diameter d$_i$ as isolated when there were no 
companions within 20$\times$d$_i$, and with size, d$_c$, between 1/4 and 
4$\times$d$_i$. To compare this with our criteria we have to assume that the 
luminosity is proportional to some power of the size, let's say L $\propto$ R$^2$. 
It is easy to verify that both criteria are similar even if Karachentseva's one is 
slightly stricter. However, Karachentseva's criterion doesn't exclude the possibility 
of small, faint perturbers that could have some effect if they are close enough to 
the primary galaxy.

\cite{mm96} defined a galaxy as isolated if there were no companions within a 
projected radius of 0.5~Mpc, and with relative redshift less than 500~km/s. 
They also made a visual inspection of the POSS images to search for faint 
companions. Applying the $f$-value criterion to their data, we find that over 
90\% of the galaxies considered isolated by \cite{mm96} are also isolated under the 
criterion here.

On their side, \cite{vsc86} considered as isolated the galaxies without companions 
with $m \le$ 14.5, within a radius which varies with the distance to the galaxy to
 keep constant the probability of finding a galaxy in a given volume. The work was 
complemented with a search for fainter companions in the POSS plates, without a
 clearly stated objective criterion. We found that 15 out of the 43 galaxies in 
their sample have faint companions. We applied our $f$-parameter selection method 
to the other 28 galaxies, to find that only 2 of them would be considered as 
isolated with our criterion. This is a clear illustration of the problems that 
can be encountered when the difference between field and isolated galaxies is not 
clearly stated.

\section{The Selection process}

The starting point of our search was the CfA catalog of galaxies (Huchra et al. 2000). 
Our aim was to build a volume limited, statistically meaningful sample of isolated 
spiral galaxies. To that end we first selected all the galaxies classified as disk 
galaxies (Spiral and Lenticular), with cz $\leq$ 5000 km/s. Only objects at high 
galactic latitudes, $|b|\geq40\deg$, were retained, to avoid problems with the 
extinction correction. Finally we considered only objects with declinations North 
of $-2.5\deg$. The number of galaxies satisfying all these conditions amounts to 
1706. The properties of these galaxies were extracted from the LEDA database 
\citep{pgc}. The distances were calculated using the LEDA kinematical distance 
modulus, corrected for extinction and Virgocentric flow.

Each galaxy was then searched for companions in the same LEDA catalog, complete to 
m = 18. The identification of the CfA objects with LEDA objects is not always 
straightforward since offsets in the coordinates up to 1\min~ are not infrequent. 
When there was a possibility of miss-identification we decided after visual 
inspection. Once all the CfA objects were identified in LEDA, we started the 
search for companions. The searching field was limited to that defined by the 
maximum distance at which a bright (massive) galaxy with $M = -23$ would produce 
a value of $f=-4.5$. Special care was taken to avoid confusion with extended 
objects that are not galaxies, and with double or distorted star images that were 
taken as galaxies in the preliminary version of the catalog we used.

We also imposed two more restrictions, on size and luminosity, to eliminate as many 
background objects as possible. The suspected companions were retained only if, 
supposed at the distance of the parent galaxy, they would be larger than 2~kpc in 
diameter, and brighter than  $M_P = -$12. For some faint galaxies present in the 
search field no magnitudes were listed in the LEDA catalog. In these cases we 
calculated the apparent magnitude that those objects would have to produce $f$ =$-$4.5,
 and we compared it with the estimation from visual 
inspection. Only companions found brighter than that value were retained as such.  
Using those restrictions we could reject most of the possible companions. For 
the doubtful cases the central galaxy was taken as perturbed. In a final step we 
discarded all the companions for which (known) redshift is different from that of 
the primary galaxy by more than 500 km/s. The $f$ values for all the possible 
perturbers within the search volume were computed, and only those with $f > -4.5$,
 and satisfying the conditions on size and luminosity were  finally taken as 
possible perturbers. The above selection process reduced the list of candidate 
galaxies to 329.

We then examined the DSS data looking for the presence of possible perturbers with 
$m_P < 18$. The process was only applied to the volume where those faint galaxies
 would still produce $f \ge -$4.5. The size and magnitude of all the possible 
perturbers were measured, and the size and luminosity criteria applied. More than 
100 galaxies were found to have companions. At the end of the process a list of 203 
isolated S and S0 galaxies was produced. This represents less than 12\% of the initial
 sample.  We insist that 
the list is not complete since all the galaxies with faint companions 
($m_P <$ 18 and $f\geq -$4.5) without known 
redshift were discarded. Besides that catalogue of isolated galaxies we 
selected another sample of non 
isolated galaxies containing systems having companions with $f > -2$, and 
$\Delta$z $<$ 500 km/s. A total of 130 objects were extracted from the original 
sample. This sample will be used as a 
comparison for the properties of isolated galaxies. In the text, we shall refer 
to this sample as {\sl perturbed} galaxies.

\section{Properties of the Isolated Disk and Perturbed Galaxies}

The database we used for the properties of the galaxies is the LEDA Catalogue. The identification and main properties of the isolated and perturbed galaxies, are presented in Table~1\footnote{Available only in electronic form} and Table~2$^1$ respectively. 

The data presented in these tables have been extracted from various sources, as described in the following. To standardize the information contained in our catalogue we extracted from the LEDA \citep{pgc} catalogue for each galaxy: PGC number, morphological type code $t$, the geometrical parameters at the 25 mag arcsec$^2$ isophote log D$_{25}$ and log r$_{25}$, the corrected colors (U-B)$_o$ and (B-V)$_o$. the mean surface brightness at the same isophote $\mu_{25}$, the kinematical parameters $W_{20}$ (the 21-cm line width at 20\% of the peak), log$\sigma$ (velocity dispersion), log$v_m$ (the maximun rotational velocity), the redshift $cz$, the kinematic distance modulus (m-M)$_{cin}$, the blue corrected absolute magnitude M$_B$, the far infrared magnitude m$_{FIR}$ and the 21-cm line magnitude m$_{21}$. The distance moduli are mainly derived from redshifts, corrected 
for Virgocentric inflow and adopting H$_0$ = 70 \kms\ Mpc$^{-1}$.  When redshift was not available, we used the photometric distance modulus, if present in LEDA. Information lacking in LEDA for some galaxies was completed using the ADS bibliographic archive, the NED database or the SIMBAD service of the Strasbourg Centre of Donn\'ee Stellaire (CDS). 

The m$_{FIR}$ and diameter values have been corrected for the distance and expresses in absolute magnitudes M$_{FIR}$ and diameters D in Kpc. The values of FIR - B presented in Table 1 represents a color index calculated as M$_{FIR} -$ M$_B$. 

Additional data on gas content were extracted from various sources in the literature, indicated in the references of Table 1. They are: 

1) the total warm dust mass, calculated from 60 $\mu$m and 100 $\mu$m IRAS fluxes (S$_{60}$ and S$_{100}$) using the the formula:

$$Log\ M_d =-2.32+Log\ S_{100}+2\ Log\ d+ Log(exp(\frac{144.06}{T_d})-1) $$ where M$_d$ is the dust mass in solar units, T$_d$ is the dust temperature in K, calculated as $ T_d=49 \times (S_{60}/S_{100})^{0.4} $. Fluxes are in mJy and distance $d$ is in Mpc.   

2) HI masses in solar units, calculated from m$_{21}$, with the expression:
 $$Log\ M_{HI} = 5.37-0.4(m_{21c}-17.4)+2\ Log\ d$$ 
or calculated from 21 -cm fluxes S$_{21}$, in Jy \kms, by the formula:
$$Log\ M_{HI} = 5.37+\ Log\ S_{21}+2\ Log\ d$$ 
 
3) molecular gas masses in solar units from CO(1--0) line fluxes ($S_{CO}$ in Jy \kms) 
by the formula: 
$$Log\ M_{mol} = 4.17 + 2\ Log\ d + Log\ S_{CO}$$
that implicitly assume a constant CO/H$_2$ conversion factor
$\chi$ = $N(H_2)/I_{CO}$= 2.3 $\times$ 10$^{20}$ mol/K km s$^{-1}$
\citep{strong}; $M_{mol}$ includes the helium mass fraction, equal to  36
the H$_2$ mass. 

When only mass values were available in the references, these have been scaled to the distances assumed here. When data for a single galaxy were available in several catalogues, we compared the mass values 
producing a weighted mean value. When both upper limits and detections were available, only detections have been used to compute mean values. Moreover, when several upper limits were available, only the lowest value has been adopted. All the above mass data were normalized by the total blue luminosity in solar units, calculated from M$_B$.

\subsection{Main properties of the samples}

\begin{figure*}
\resizebox{9cm}{!}{\includegraphics[angle=-90]{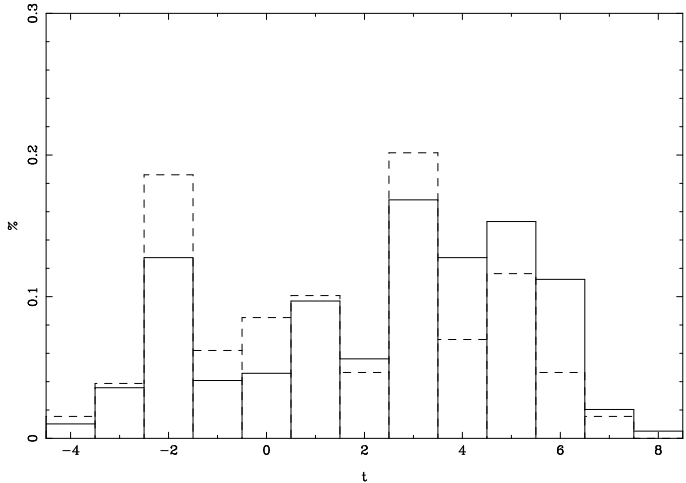}} 
\caption{The normalized distributions of morphological types for Isolated ({\it 
solid line}) and Perturbed ({\it dashed line}) galaxies.} 
\label{tipos}
\end{figure*}

The first aspect we considered was the distribution of the galaxies in both samples 
by  morphological types (Figure~\ref{tipos}). The differences between the isolated 
and perturbed galaxy distributions are apparent. Given the clear non-normality of the 
distributions, we applied the Mann-Whitney U-test. The results indicate that the 
two distributions are different with a significance level $>$ 99.5\%. The largest 
differences are
found for Sc galaxies, that are more abundant among isolated galaxies (in agreement 
with earlier results by M\'arquez 1994, M\'arquez \& Moles, 1996, and M\'arquez 
et al., 1999), and for S0 galaxies, that are more abundant among perturbed galaxies 
(see also Table~ 3).
%
\noindent

For the purpose of comparing all the other properties of isolated and perturbed 
galaxies we  grouped them into three morphological categories, S0 
(-4.5 $<$ t $\leq$ -0.5), early Sp (-0.5 $<$ t $\leq$ 3.5) and late Sp (t $>$ 3.5).

Regarding the frequency of bars, it appears that barred galaxies are far more 
abundant (about twice) among perturbed galaxies. The difference is very marked 
for Lenticular and early Sp types, whereas there is no difference for late Sp 
galaxies (see Table~3). We also compared all the other catalogued properties of 
the two samples. The differences were tested with the Mann-Whitney U-test. We 
found that the two samples are different at a highly significant level ($>$ 99.5\%) 
in all the main catalogued properties, including the absolute blue 
magnitude, M$_B$, the infrared luminosity normalized to the B luminosity, 
$M_{FIR} -$ M$_B$, the size, D$_{25}$, the color indexes (U$-$B) and (B$-$V), 
the mean surface brightness within the isophote 25, $\mu_{25}$, the maximum 
velocity rotation of the gas, v$_m$, the stellar central velocity dispersion, 
$\sigma$, and the 21-cm line width at 20\% of the peak, W$_{20}$. In Table~3 
we give the median values of those properties for the different families. 

The only property that appears to be similar for both samples is the amount of 
atomic gas as measured by $\log(M_{HI}/L_B)$. Looking at the different morphological 
bin there is a hint on a possible difference for lenticular galaxies.

It can be seen that the isolated galaxies are smaller, less luminous and 
bluer than the perturbed systems. We also notice the consistency of the results 
regarding the dynamical variables. Indeed, V$_{max}$, $\sigma$ and W$_{20}$ are 
smaller for the isolated galaxies. They also present lower FIR luminosity and 
molecular gas content. Notice that the same trend is seen in all the morphological
 bins we considered, even if the largest and more significant differences are found 
in most cases for the lenticular and early Sp galaxies. 

Since the content of molecular gas is not given in LEDA we searched the literature 
for it, using the same references and approach as Bettoni et al. (2003; the 
references are given in the Notes of Tables 1 and 2). We found data for only 21 
isolated and 20 perturbed galaxies. We found that the perturbed galaxies have 
higher M$_{mol}/L_B$ values, the difference being statistically significant at 
$>$ 99.5\%. Notice that the difference arise only in the late SP morphological bin. 

\begin{table*}
\setcounter{table}{2}
\caption[]{Comparison of the main properties of Isolated (Is) and Perturbed (Pt) 
galaxies. The statistical parameter z from the U-test is given in the third column. 
In the last three columns we give the median values for the three morphological 
bins defined in the text. The first row contains the results for the morphological 
types, the numbers in parentheses corresponding to the number of galaxies used in 
the comparison. The second row gives the fraction of galaxies with bars. The other
 rows correspond to the other properties we considered.} 
      \tabcolsep 0.1truecm
      \center
            \begin{tabular}{|l|l|r|r|r|r|}
      \hline
          & Status &  z-value & S0 & Early Sp &  Late Sp \\
        \hline

Type     & Is (196) & 2.79 & 21.4\%  & 36.7\% & 41.8\% \\
         & Pt (129) &      & 31.0\%  & 43.4\% & 25.6\% \\
\hline
Bars    &  Is (64) & -- & 15.7\% & 34.9\% & 38.2\% \\
        &  Pt (66) &    & 39.2\% & 71.1\% & 41.2\% \\
\hline
M$_B$       & Is (203) & 4.74 & $-$17.88 (42) & $-$19.55 (72) & $-$19.27 (89)  \\
            & Pt (130) &      & $-$19.60 (40) & $-$19.90 (56) & $-$19.73 (34)  \\
         \hline
log D$_{25}$ & Is (203) & 6.05 & 0.90 (42) & 1.21 (72) & 1.20 (89)  \\
             & Pt (130) &      & 1.28 (40) & 1.38 (56) & 1.35 (34)  \\
         \hline
(U-B)  & Is (45) &  3.84 & 0.22 (11) & 0.03 (19) & $-$0.12 (15)  \\
       & Pt (67) &       & 0.40 (26) & 0.15 (27) & $-$0.07 (14)  \\
   \hline
(B-V) & Is (58) &  3.69  & 0.76 (11) & 0.66 (19) & 0.51 (28)  \\
      & Pt (78) &        & 0.85 (29) & 0.68 (31) & 0.53 (18)  \\
 \hline
$\mu_{25}$ & Is (203) & 2.73 & 23.29 (42) & 23.04 (72) & 23.39 (89)  \\
           & Pt (130) &      & 23.50 (40) & 23.38 (56) & 23.38 (34)  \\

 \hline
log v$_m$ & Is (149) & 3.99 & 1.87 (13) & 2.20 (55) & 2.09 (81)  \\
          & Pt  (94) &      & 2.21 (15) & 2.31 (48) & 2.13 (31)  \\
 \hline
log $\sigma$ & Is (23) & 3.15 & 2.06 (10) & 2.23  (6) & 1.99 (7)  \\
             & Pt (46) &      & 2.25 (26) & 2.18 (16) & 2.05 (4) \\
\hline
W$_{20}$ &  Is (139) &  5.09 & 164 (13) & 290 (53) & 240 (73)  \\
         &  Pt (87)  &       & 327 (14) & 386 (45) & 264 (28)  \\
\hline
M$_{FIR} - $M$_B$ & Is (104) & 3.02 & $-$0.80 (11) & $-$0.80 (45) & $-$0.31 (48) \\
                  & Pt  (44) &      & $-$1.06  (1) & $-$0.34 (27) & $-$0.25 (16) \\
\hline   
log($M_{HI}/L_B$) & Is (151) & 0.91 & $-$0.74 (14) & $-$0.67 (55) & $-$0.46 (82) \\  
                  & Pt (93)  &      & $-$1.17 (15) & $-$0.66 (47) & $-$0.38 (31) \\
\hline
log($M_{mol}/L_B$) & Is (21) & 2.61 & $-$0.89 (2)  & $-$0.64  (5)  & $-$1.15 (14) \\
                  & Pt (20)  &      & --           & $-$0.62 (13)  & $-$0.62 (7) \\
\hline
        \end{tabular}
\end{table*}

\begin{figure}
\resizebox{9cm}{!}{\includegraphics[angle=-90]{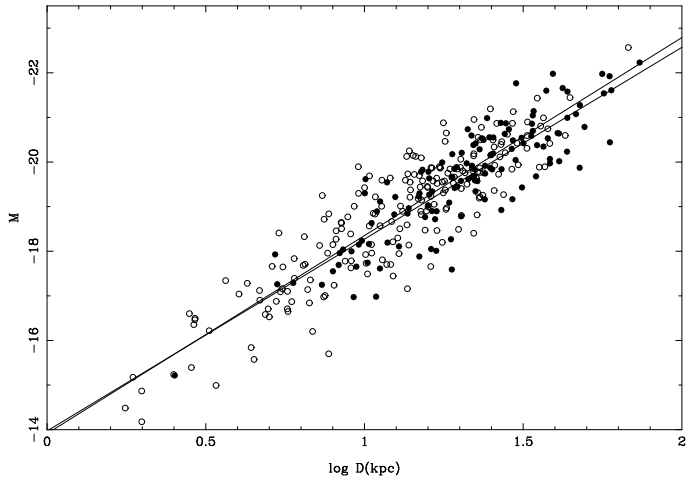}} 
\caption{The Luminosity-size relation for Isolated ({\it empty circles}) and 
Perturbed ({\it filled circles}) galaxies. The lines are the best fit to the two 
families} 
\label{Lsize}
\end{figure}

We analyzed the relation between global structural parameters as well. We found that
 both families satisfy the same Luminosity-Size relation (Figure~\ref{Lsize}). 
We notice the almost absence of bright and big isolated galaxies, together with 
the almost absence of small and faint perturbed galaxies. This tendency cannot be 
due to the selection criteria, based on $f$ values, and depending on sizes and 
luminosity by means of equation \ref{f}. In fact, more massive and less extended 
systems are less subject to perturbations by the surrounding galaxies. One may 
expect that galaxies selected on the basis of a lower $f$ could be biased toward 
smaller but brighter systems. On the contrary, isolated galaxies in our sample 
appear smaller, i.e., producing smaller $f$-values, but fainter, i.e., producing 
higher $f$-values.

\begin{figure}
\resizebox{9cm}{!}{\includegraphics[angle=-90]{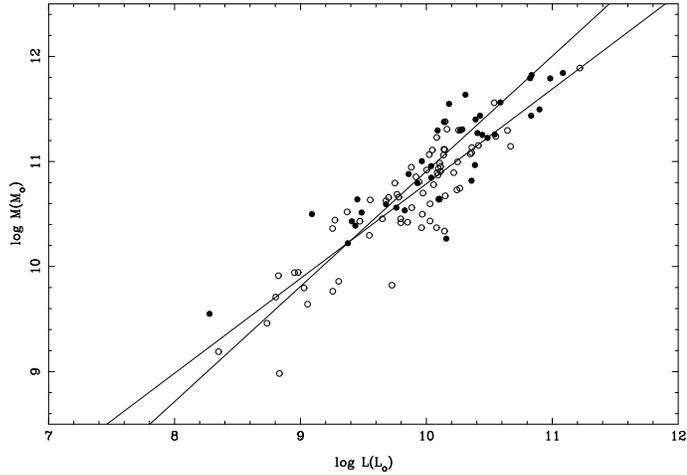}} 
\caption{The Mass-Luminosity relation for Isolated and Perturbed galaxies. Same 
symbols as in Fig 3. The rotation velocity was corrected for inclination. Only 
galaxies not later than Scd, with inclinations between 40 and 70 degrees have 
been plotted. The lines are the fits to both families} 
\label{ML}
\end{figure}

In Figure~\ref{ML} we present the M {\it versus} L relation. The masses have been 
estimated using a central, point like mass model, with M(M$_{\odot}$) = 
2.3265$\times$v$^2 R_{25}$ (v, from LEDA, in km/s and $R_{25}$ in kpc). It can be seen 
in the Figure that both families, isolated and perturbed galaxies, define very 
approximately the same relation, and share an important region in the diagram. 
The differences arise because there are essentially no isolated galaxies with 
high mass (and luminosity), whereas there are no perturbed galaxies with low mass 
(and luminosity).

\noindent

\begin{figure}
\resizebox{9cm}{!}{\includegraphics[angle=-90]{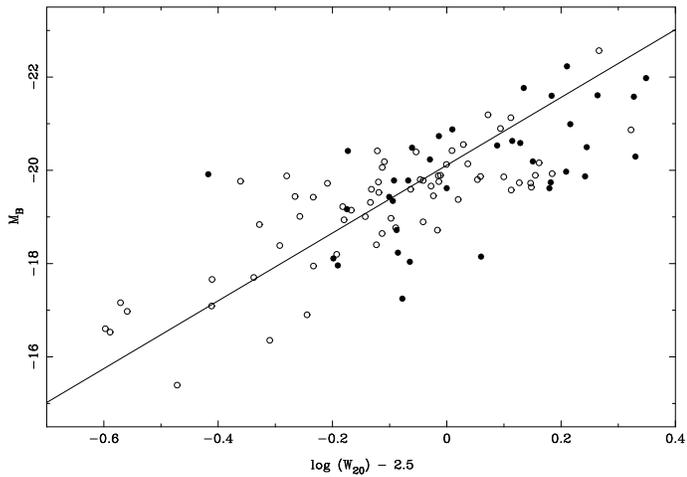}} 
\caption{The Tully-Fisher relation for Isolated and Perturbed galaxies. Same 
symbols and galaxies as in Figure 4. The line represents the fit to the T-F 
relation as given by Tully and Pierce (2002).}
\label{TF}
\end{figure}

The Tully-Fisher relation for S galaxies in both samples is presented in 
Figure~\ref{TF}. Even if both families follow the standard TF relation 
(Tully \& Pierce, 2000), the scatter is important and no conclusion about 
possible differences can be extracted before more homogeneous and accurate 
data are available. In any case, the presence of bars do not have, apparently, 
any influence on the position of a galaxy in the T-F diagram, in agreement with 
the results reported by Courteau et al. (2003).

\noindent
%

\section{Conclusions}

We used a well defined, physically meaningful criterion to define isolated galaxies.
 We performed a search of isolated galaxies using as the parameter the logarithmic 
ratio $f$ between the internal and tidal forces at the outskirts of a given galaxy.
 The adopted limit, $f < -$4.5, was checked with results from numerical simulations 
\citep{byrd92}, and with data from the Coma Cluster. Similarly, a sample of {\sl
 perturbed} galaxies was defined comprising of galaxies with confirmed companions 
satisfying $f \geq -$2.

Comparison of the properties of the galaxies in both samples was made using the 
Mann-Whitney U-test. The first result to notice is the significant differences 
in morphological types, with Sc types being more abundant among isolated galaxies,
 whereas S0 galaxies are significantly more abundant among perturbed systems. 
We also found that barred galaxies are more frequent (about twice) among perturbed 
than among isolated galaxies except for late type Spiral galaxies.

They also differ in all the properties we have examined, except in atomic gas 
content, with similar values of the ratio M$_{HI}$/L$_B$. The isolated galaxies 
appear to be smaller and fainter, with bluer color indices. They have less 
molecular gas an FIR luminosity per unit of blue luminosity as well. The 
dynamical parameters are also consistently smaller for isolated galaxies. 
Even if the differences can be appreciable for all the families (S0, early 
Sp and late Sp), the biggest differences are for S0 and early Sp. The exception 
is in molecular gas content, for which the maximum difference is found for 
late type Sp galaxies.

The M-L and M-size relations are consistent in showing the absence of big, 
luminous and massive systems among the isolated galaxies. Moreover, we also 
found the almost complete absence of perturbed systems with small sizes, low 
luminosities and low masses. We also found that barred galaxies do not occupy 
a particular region in the T-F diagram.

In view of all these results it is tempting to consider the differences as arising
 from the different evolving conditions. Our results indicate that the gravitational 
interaction in aggregates would produce evolution from late Spiral, relatively 
faint and low mass galaxies, toward earlier, more luminous and massive Spiral
 and Lenticular types. This would also favor the formation of bars in early type 
Spirals and Lenticulars. However, the relations between global parameters are 
similar for both families, even if they tend to occupy different regions in the 
corresponding plots (see Figures 2, 3 and 4)

Since the isolated galaxies wouldn't have had the opportunity to accrete other 
systems and grow in the way the hierarchical models predict, they would still 
be similar today to the high redshift systems supposed to be the {\sl building 
blocks} for the formation of larger galaxies. This view is consistent with 
the results reported by \citet{ferg03}, showing that the sizes of galaxies at 
z $\approx$ 4 are smaller than nearby luminous galaxies. If cannibalism and 
accretion in the early phases of galaxy evolution are important to fix the final 
global properties such as size, luminosity and mass, isolated galaxies would 
resemble those initial pieces, the left over fragments in the early process of 
galaxy formation.


\begin{acknowledgements}
JV, MM, IM and JM acknowledge financial support from the Spanish Ministerio de 
Ciencia y Tecnolog\'{\i}a through grants PB98-0684, AYA2002-01241 and AYA2001-2089, 
and from the Junta de Andaluc\'{\i}a, grant TIC-144. GG has made use of funds from 
the University of Padova (Fondi 60\% - 2002). We thank Dr. G. Paturel for allowing 
us to use the new LEDA catalogue prior to publication. 

\end{acknowledgements}


%
\newpage
\begin{landscape}
\setcounter{table}{0}
\begin{table}
      \caption[]{Main properties of the Isolated galaxies. All the propertis were extracted from LEDA except the molecular gas content}
\tabcolsep 5pt
            \begin{tabular}{lrrrrrrrrrrrrrrrll}
PGC	&      t &   log D$_{25}$ 	& log r$_{25}$ & (U-B)$_o$	& (B-V)$_o$ & $\mu_{25}$ 	& W$_{20}$ 	& log $\sigma$ & log v$_m$  &	cz &	(m-M)$_{cin}$ & M$_B$ & FIR-B & log $\frac{M_{HI}}{L_B}$ & log $\frac{M_{mol}}{L_B}$ & Ref.	&	Notes	\\
	&        &   0.'1 	&   0.'1 &                &           & mag arcsec$^{2}$ 	& km s$^{-1}$ 	&  km s$^{-1}$  & km s$^{-1}$ & km s$^{-1}$ &	      &       &       &                          &                           &  	&		\\

\hline
218	&	2.0	&	1.75	&	0.33	&	0.40	&	0.88	&	23.68	&	460	&	2.23	&	2.38	&	1053	&	31.01	&	-19.93	&		&		-1.04	&	$<$	-1.11	&	b,d	&		\\
1525	&	5.9	&	1.73	&	0.93	&	-0.15	&	0.41	&	24.49	&	221	&		&	1.99	&	842	&	30.54	&	-18.49	&		&		0.02	&			&	d	&		\\
2600	&	5.3	&	1.20	&	0.07	&		&		&	23.29	&		&		&	2.24	&	4452	&	34.04	&	-20.87	&	-1.51	&		-0.79	&			&	d	&		\\
3043	&		&	0.87	&	0.22	&		&		&	23.44	&	192	&		&	1.98	&	1623	&	31.77	&	-16.58	&		&		0.46	&			&	d	&		\\
3763	&	2.2	&	0.94	&	0.31	&		&		&	22.91	&	181	&		&	1.93	&	4668	&	34.15	&	-19.88	&	-2.56	&		-0.77	&			&	d	&		\\
4785	&	-0.1	&	1.29	&	0.22	&	0.06	&	0.66	&	23.47	&	244	&		&	2.08	&	2133	&	32.46	&	-19.55	&	0.03	&		-0.83	&			&	c,d	&		\\
4948	&	4.1	&	1.24	&	0.67	&		&		&	23.23	&	381	&		&	2.23	&	2509	&	32.81	&	-19.85	&	-2.08	&		-0.21	&			&	d	&		\\
5139	&	5.3	&	1.56	&	0.08	&		&	0.52	&	23.72	&	266	&		&	2.30	&	2469	&	32.75	&	-20.75	&	0.72	&		-0.57	&			&	d	&		\\
5194	&	5.7	&	1.18	&	0.82	&		&		&	24.13	&	161	&		&	1.81	&	2415	&	32.72	&	-18.47	&		&		-0.09	&			&	d	&		\\
5232	&	-2.0	&	1.18	&	0.35	&		&		&	24.49	&		&		&		&	1930	&	32.20	&	-17.70	&		&			&			&		&		\\
5321	&	5.9	&	1.07	&	0.16	&		&		&	23.28	&	262	&		&	2.18	&	4129	&	33.87	&	-20.06	&		&		-0.76	&			&	d	&		\\
5634	&	5.3	&	1.20	&	0.04	&		&		&	23.80	&	153	&		&	2.17	&	3148	&	33.30	&	-19.50	&		&		-0.47	&			&	d	&		\\
5643	&	3.0	&	1.41	&	0.73	&		&		&	24.23	&	395	&		&	2.26	&	2805	&	33.04	&	-19.94	&	-0.60	&		-0.51	&			&	d	&		\\
5998	&	5.9	&	0.88	&	0.18	&		&		&	22.10	&	212	&		&	2.07	&	3183	&	33.27	&	-19.43	&		&		-0.85	&			&	d	&		\\
6275	&	3.1	&	1.39	&	0.24	&	-0.18	&	0.55	&	23.57	&	322	&		&	2.23	&	2987	&	33.16	&	-20.42	&	-0.41	&		-0.27	&			&	d	&		\\
6656	&	0.0	&	1.58	&	0.53	&		&		&	24.21	&	397	&		&	2.26	&	1508	&	31.63	&	-19.26	&		&		-0.41	&			&	c,d	&		\\
6893	&	5.4	&	1.06	&	0.19	&		&		&	24.11	&	250	&		&	2.13	&	4701	&	34.15	&	-19.31	&		&		0.12	&			&	d	&		\\
6897	&	4.6	&	1.23	&	0.09	&	-0.04	&	0.59	&	22.97	&	313	&		&	2.35	&	4984	&	34.25	&	-21.43	&	-1.17	&		-0.97	&			&	d	&		\\
6993	&	1.1	&	1.37	&	0.04	&	0.33	&	0.83	&	23.10	&	124	&	2.13	&	2.10	&	1728	&	31.92	&	-19.58	&	1.28	&		-2.58	&			&	d,e	&		\\
7577	&	3.7	&	1.18	&	0.63	&		&		&	22.37	&		&		&		&	3486	&	33.49	&	-20.95	&		&			&			&		&		\\
7826	&	5.4	&	0.96	&	0.04	&		&		&	23.76	&	96	&		&	1.97	&	2379	&	32.69	&	-17.79	&		&		0.11	&			&	d	&		\\
7952	&	6.6	&	1.21	&	0.06	&		&		&	24.57	&		&		&	1.96	&	3410	&	33.42	&	-18.82	&		&		0.13	&			&	d	&		\\
8109	&		&	0.93	&	0.35	&		&		&	23.50	&	248	&		&	2.06	&	4516	&	34.06	&	-19.23	&	-1.79	&		-0.36	&			&	d	&		\\
8163	&	3.3	&	0.99	&	0.32	&		&		&	22.31	&		&		&		&	4410	&	34.00	&	-20.65	&	-0.79	&			&			&	d	&		\\
8165	&	3.0	&	1.13	&	0.25	&		&		&	22.47	&	363	&		&	2.29	&	4419	&	34.01	&	-21.19	&	-0.55	&		-0.61	&			&	d	&		\\
9126	&	6.5	&	1.20	&	0.14	&		&		&	23.45	&	118	&		&	1.83	&	1385	&	31.36	&	-17.78	&		&		-0.86	&			&	d	&		\\
9988	&	1.0	&	1.38	&	0.05	&		&		&	24.27	&	181	&		&	2.19	&	2630	&	32.81	&	-19.36	&		&		-0.17	&			&	d	&		\\
10789	&	-1.8	&	1.00	&	0.39	&		&		&	22.08	&		&		&		&	2596	&	32.78	&	-19.69	&		&			&			&		&		\\
10815	&	6.5	&	1.16	&	0.33	&		&		&	23.76	&		&		&	2.17	&	4502	&	34.01	&	-20.12	&		&		-0.38	&			&	d	&		\\
10942	&	3.0	&	1.25	&	0.24	&		&		&	23.77	&	290	&		&	2.18	&	3040	&	33.15	&	-19.80	&	-0.83	&		-0.11	&			&	d	&		\\
\hline 
        \end{tabular}
\end{table}
\clearpage

\begin{table}
            \begin{tabular}{lrrrrrrrrrrrrrrrll}
\hline 
25467	&	0.1	&	1.04	&	0.28	&		&		&	23.05	&		&		&		&	2992	&	33.22	&	-19.21	&	-0.31	&			&			&	d	&		\\
25985	&	2.2	&	0.89	&	0.22	&		&		&	23.45	&		&		&		&	1934	&	32.33	&	-17.14	&		&			&			&		&		\\
26218	&	-2.0	&	1.14	&	0.27	&		&		&	23.65	&		&		&		&	1660	&	31.98	&	-17.93	&	-0.95	&			&			&	c,d	&		\\
26512	&	3.0	&	1.88	&	0.32	&	0.27	&	0.79	&	22.64	&	607	&	2.33	&	2.50	&	638	&	30.38	&	-20.87	&	1.36	&		-0.87	&		-0.94	&	b,d	&	CR	\\
26690	&	4.1	&	1.41	&	0.69	&		&		&	23.23	&	601	&		&	2.46	&	4097	&	33.89	&	-21.45	&		&		-0.37	&			&	d	&		\\
26979	&	0.5	&	1.03	&	0.07	&	-0.08	&	0.62	&	22.88	&	172	&		&	1.98	&	1697	&	32.06	&	-18.15	&	-1.70	&		-0.57	&			&	d	&		\\
27077	&	4.0	&	2.10	&	0.36	&	-0.02	&	0.57	&	23.16	&	384	&	2.01	&	2.37	&	556	&	29.76	&	-20.90	&	-0.24	&		-0.94	&		-1.02	&	b,d,f	&		\\
27157	&	-1.0	&	0.64	&	0.15	&		&		&	22.32	&		&		&		&	1473	&	31.80	&	-16.50	&		&			&			&		&		\\
27311	&	2.8	&	0.89	&	0.43	&		&		&	23.50	&		&		&		&	1654	&	32.00	&	-16.71	&		&			&			&		&		\\
27437	&	1.1	&	1.14	&	0.31	&		&		&	23.62	&	308	&		&	2.18	&	4060	&	33.87	&	-19.76	&		&		-0.51	&			&	d	&		\\
27518	&	4.0	&	1.15	&	0.23	&		&		&	22.81	&	390	&		&	2.34	&	4981	&	34.26	&	-21.13	&	-0.46	&		-0.79	&			&	d	&		\\
27792	&	4.3	&	1.00	&	0.09	&		&		&	23.29	&		&		&	1.97	&	1466	&	31.83	&	-17.36	&		&		0.09	&			&	d	&		\\
27796	&	3.1	&	1.24	&	0.46	&		&		&	23.45	&		&		&	2.34	&	4940	&	34.28	&	-20.81	&	-0.05	&		-0.44	&			&	d	&		\\
27968	&	6.5	&	1.10	&	0.30	&		&		&	24.89	&	191	&		&	1.95	&	3088	&	33.22	&	-17.66	&		&		0.42	&			&	d	&		\\
28145	&	1.5	&	0.94	&	0.47	&		&		&	22.76	&		&		&		&	4679	&	34.19	&	-19.86	&	-1.90	&			&			&	d	&		\\
28259	&	-1.7	&	0.92	&	0.07	&		&		&	23.21	&	198	&		&	2.14	&	1524	&	31.86	&	-17.10	&		&		-1.11	&			&	d	&	CR	\\
28401	&	3.6	&	1.13	&	0.04	&		&		&	23.75	&	60	&		&	1.55	&	3365	&	33.56	&	-19.88	&	-0.48	&		-0.41	&			&	d	&		\\
28424	&	-1.9	&	1.25	&	0.04	&	0.09	&	0.63	&	23.29	&	241	&	1.91	&	2.28	&	1538	&	31.86	&	-18.67	&	-0.81	&		-1.43	&		-0.75	&	b,c,d,e	&		\\
28485	&	5.3	&	1.57	&	0.19	&		&	0.69	&	23.54	&	294	&		&	2.19	&	1412	&	31.60	&	-19.78	&	0.38	&		-0.73	&			&	d	&		\\
28672	&	3.1	&	1.24	&	0.50	&		&		&	22.45	&	312	&		&	2.12	&	2986	&	33.31	&	-20.78	&	0.12	&		-1.14	&			&	d	&		\\
28758	&	0.8	&	0.84	&	0.22	&		&		&	22.91	&	202	&		&	1.93	&	1486	&	31.82	&	-16.90	&		&		-0.24	&			&	d	&		\\
29009	&	1.1	&	1.25	&	0.15	&		&		&	23.11	&	261	&		&	2.18	&	2406	&	32.73	&	-19.75	&	-2.31	&		-0.63	&			&	d	&		\\
29177	&	-3.1	&	0.74	&	0.07	&	-0.27	&	0.53	&	22.25	&		&		&		&	2605	&	33.04	&	-18.33	&	-1.68	&		-0.27	&			&	d	&		\\
29198	&		&	0.63	&	0.19	&		&		&	23.15	&		&		&		&	1112	&	31.02	&	-14.87	&		&			&			&		&		\\
29347	&	-2.4	&	0.82	&	0.03	&	-0.14	&	0.39	&	22.54	&	115	&		&	2.03	&	1362	&	31.62	&	-17.04	&		&		-0.97	&			&	d	&		\\
29715	&	3.0	&	1.26	&	0.46	&		&		&	24.52	&	442	&		&	2.33	&	4754	&	34.21	&	-19.89	&	-0.95	&		0.01	&			&	d	&		\\
30010	&	3.0	&	1.19	&	0.12	&	0.03	&	0.59	&	22.47	&	276	&		&	2.19	&	1308	&	31.49	&	-18.77	&	-2.03	&		-1.06	&			&	d	&		\\
30197	&	5.2	&	1.89	&	0.48	&	-0.12	&	0.43	&	23.39	&	322	&	1.81	&	2.19	&	663	&	30.43	&	-20.21	&	0.69	&		-0.37	&			&	d	&		\\
30310	&	4.0	&	0.92	&	0.13	&		&		&	24.18	&		&		&	1.93	&	2906	&	33.12	&	-17.49	&		&		0.13	&			&	d	&		\\
30569	&	5.9	&	1.27	&	0.32	&		&		&	23.73	&	259	&		&	2.11	&	2127	&	32.64	&	-18.97	&	-0.31	&		-0.09	&			&	d	&		\\
30858	&	4.5	&	0.90	&	0.15	&		&		&	23.59	&		&		&		&	2502	&	32.96	&	-17.63	&		&			&			&		&		\\
30895	&	4.0	&	1.63	&	0.40	&	0.01	&	0.58	&	23.82	&	431	&	2.06	&	2.31	&	1352	&	31.62	&	-19.73	&		&		-0.44	&			&	d	&		\\
31304	&	-0.9	&	1.08	&	0.11	&		&		&	22.89	&		&		&		&	957	&	30.63	&	-17.12	&		&			&			&		&		\\
\hline 
        \end{tabular}
\end{table}
\clearpage

\begin{table}
            \begin{tabular}{lrrrrrrrrrrrrrrrll}
\hline 
31472	&	-1.9	&	1.21	&	0.28	&		&		&	23.42	&		&	2.20	&		&	3054	&	33.24	&	-19.76	&		&	$<$	-1.57	&			&	c,e	&		\\
31601	&		&	0.65	&	0.00	&		&		&	21.95	&	118	&		&		&	1699	&	32.22	&	-17.34	&		&		-0.61	&			&	d	&		\\
31650	&	4.0	&	1.44	&	0.02	&	-0.45	&	0.32	&	21.98	&	233	&	2.05	&	2.24	&	988	&	31.19	&	-20.25	&	-1.45	&		-0.63	&		-1.50	&	b,d,g	&	ARP217	\\
31883	&	5.1	&	1.61	&	0.22	&	-0.07	&	0.51	&	23.47	&	352	&		&	2.26	&	1301	&	31.44	&	-19.87	&	0.08	&		-0.13	&		-1.11	&	b,d	&		\\
31945	&	-2.0	&	1.11	&	0.49	&	-0.14	&	0.60	&	23.86	&	121	&		&	1.63	&	1321	&	31.52	&	-17.01	&	-0.80	&		-0.53	&			&	d	&		\\
32183	&	5.2	&	1.86	&	0.20	&	-0.24	&	0.41	&	23.92	&	255	&		&	2.14	&	1012	&	31.28	&	-20.42	&	0.40	&		-0.18	&		-1.21	&	b,d	&		\\
32364	&	0.1	&	0.98	&	0.36	&		&		&	22.43	&	167	&		&	1.82	&	712	&	30.12	&	-16.36	&		&		-0.81	&			&	d	&		\\
32543	&	-1.8	&	1.26	&	0.36	&		&		&	22.98	&	186	&		&	1.86	&	646	&	30.27	&	-17.39	&	0.21	&		-0.20	&	$<$	-0.82	&	b,c,d	&		\\
32719	&	4.0	&	1.48	&	0.38	&		&		&	23.54	&	298	&		&	2.15	&	1258	&	31.70	&	-19.66	&	0.56	&		-0.32	&			&	d	&		\\
33140	&	3.1	&	1.37	&	0.03	&		&		&	23.18	&	142	&		&	2.16	&	1434	&	31.66	&	-19.18	&		&		0.07	&			&	d	&		\\
33375	&	0.0	&	0.87	&	0.06	&		&		&	22.53	&		&		&		&	1549	&	32.03	&	-17.65	&		&			&			&		&		\\
33604	&	0.1	&	1.16	&	0.52	&		&		&	22.41	&		&		&		&	1344	&	31.50	&	-18.63	&		&			&			&		&		\\
33726	&		&	0.98	&	0.07	&		&		&	23.61	&	100	&		&	1.84	&	1228	&	31.59	&	-16.65	&		&		0.23	&			&	d,g	&		\\
34353	&	1.1	&	1.56	&	0.59	&		&		&	24.91	&	214	&		&	1.95	&	719	&	30.57	&	-17.16	&	-0.60	&		0.57	&			&	d	&		\\
34692	&	4.4	&	1.07	&	0.40	&		&	0.39	&	22.01	&	158	&		&	1.83	&	1314	&	31.76	&	-18.84	&	-0.04	&		-0.56	&			&	d	&		\\
34767	&	5.2	&	1.69	&	0.08	&		&	0.54	&	23.11	&	124	&		&	1.92	&	1159	&	31.49	&	-20.62	&	-0.17	&		-0.68	&		-0.55	&	b,d	&	ARP27	\\
34836	&	4.6	&	1.57	&	0.28	&	-0.10	&	0.55	&	23.08	&	532	&		&	2.54	&	4256	&	33.97	&	-22.57	&		&		-0.72	&			&	d	&		\\
34908	&	-2.0	&	1.08	&	0.21	&		&		&	23.19	&		&		&		&	2050	&	32.52	&	-18.57	&		&			&			&	c	&		\\
34935	&	4.9	&	1.17	&	0.18	&	0.01	&	0.58	&	21.51	&	310	&		&	2.26	&	1480	&	31.75	&	-19.89	&	-1.07	&		-0.99	&			&	d	&		\\
34967	&	7.8	&	0.95	&	0.53	&		&		&	22.13	&	269	&		&	2.06	&	2607	&	32.86	&	-19.30	&	-0.75	&		-0.45	&			&	d	&		\\
35025	&	4.7	&	1.17	&	0.24	&		&		&	22.71	&	242	&		&	2.09	&	1570	&	32.10	&	-19.01	&	0.11	&		-1.00	&			&	d	&		\\
35164	&	3.0	&	1.80	&	0.25	&		&		&	23.30	&	429	&	2.08	&	2.37	&	767	&	30.70	&	-20.16	&		&		-0.87	&		-0.64	&	b,d	&		\\
35225	&		&	1.04	&	0.38	&		&		&	23.43	&		&		&	1.30	&	1039	&	30.96	&	-16.71	&		&		-0.80	&			&	d	&		\\
35266	&	0.0	&	1.25	&	0.17	&	0.01	&	0.71	&	23.16	&	295	&		&	2.21	&	1512	&	32.02	&	-18.90	&	-1.46	&		-1.17	&			&	c,d	&		\\
35314	&	3.1	&	1.50	&	0.66	&		&		&	22.77	&	404	&		&	2.28	&	1724	&	32.02	&	-20.50	&	1.22	&		-0.82	&			&	d	&		\\
35440	&	2.4	&	1.66	&	0.39	&	0.04	&	0.67	&	23.30	&	354	&		&	2.23	&	1017	&	30.92	&	-19.80	&	0.35	&		-0.63	&			&	d	&		\\
35608	&	2.1	&	1.22	&	0.38	&		&		&	23.33	&	223	&		&	1.99	&	1913	&	32.48	&	-19.15	&	-0.06	&		-1.00	&			&	d	&		\\
35676	&	5.1	&	1.78	&	0.17	&		&	0.44	&	23.22	&	288	&	1.86	&	2.16	&	857	&	30.92	&	-20.39	&	0.81	&		-0.53	&		-0.95	&	b,d	&		\\
35955	&	-2.0	&	1.06	&	0.02	&		&		&	22.68	&		&		&		&	1269	&	31.68	&	-18.13	&		&			&			&	c	&		\\
36037	&	-3.3	&	0.99	&	0.00	&		&		&	22.88	&		&		&		&	1337	&	31.78	&	-17.71	&		&			&			&	c	&		\\
36043	&	-2.0	&	1.08	&	0.10	&		&		&	23.28	&	146	&		&	1.87	&	976	&	30.87	&	-16.88	&	-1.14	&		-0.80	&			&	c,d	&		\\
36211	&	-3.8	&	0.86	&	0.28	&		&		&	23.61	&	164	&		&	1.81	&	1837	&	32.26	&	-16.87	&		&		-0.10	&			&	d	&		\\
36215	&	3.8	&	1.11	&	0.15	&		&		&	22.55	&		&		&		&	1255	&	31.66	&	-18.46	&		&			&			&		&		\\
\hline 
        \end{tabular}
\end{table}
\clearpage

\begin{table}
            \begin{tabular}{lrrrrrrrrrrrrrrrll}
\hline 
36266	&	3.3	&	1.30	&	0.29	&	-0.14	&	0.50	&	21.99	&	316	&		&	2.20	&	1466	&	31.85	&	-20.13	&		&		-0.67	&			&	d	&		\\
36686	&	-2.0	&	1.02	&	0.09	&	-0.24	&	0.43	&	22.37	&	123	&		&	1.83	&	755	&	30.73	&	-17.28	&	-0.56	&		-0.67	&			&	c,d	&		\\
36776	&	3.2	&	1.01	&	0.00	&		&		&	22.79	&	172	&		&	3.10	&	3572	&	33.69	&	-19.88	&		&		-1.13	&			&	d	&		\\
36930	&	5.8	&	1.42	&	0.78	&		&		&	23.83	&	177	&		&	1.86	&	848	&	30.86	&	-17.81	&		&		-0.24	&			&	d	&		\\
37213	&	5.0	&	0.82	&	0.10	&		&	0.20	&	22.35	&		&		&	1.68	&	1055	&	30.90	&	-16.46	&		&		-0.29	&			&	d	&		\\
37235	&	-2.0	&	1.55	&	0.15	&	0.41	&	0.87	&	22.71	&	258	&	2.12	&	2.13	&	921	&	30.99	&	-19.87	&		&		-1.40	&			&	d,e	&	CR	\\
37244	&	-3.1	&	1.05	&	0.13	&		&		&	23.49	&		&	2.05	&		&	3650	&	33.67	&	-19.31	&		&			&			&		&		\\
37290	&	4.0	&	1.43	&	0.23	&		&	0.38	&	22.15	&	281	&		&	2.13	&	800	&	30.82	&	-19.59	&	-0.78	&		-0.72	&		-1.18	&	b,d	&		\\
37352	&	7.1	&	1.20	&	0.50	&		&		&	23.00	&	246	&		&	2.02	&	2384	&	32.67	&	-19.43	&		&		-0.50	&			&	d	&		\\
37444	&	5.9	&	1.40	&	0.06	&		&		&	24.23	&	187	&		&	2.20	&	1892	&	32.19	&	-18.80	&		&		-0.10	&			&	d	&		\\
37574	&	-2.0	&	1.02	&	0.10	&		&		&	23.61	&		&		&		&	3309	&	33.48	&	-18.82	&		&			&			&		&		\\
37584	&	6.7	&	1.52	&	0.08	&		&		&	24.09	&	132	&		&	1.96	&	778	&	30.78	&	-18.14	&		&		-0.16	&			&	d	&		\\
37795	&	-2.4	&	0.90	&	0.00	&		&		&	22.90	&		&		&		&	3139	&	33.37	&	-18.83	&		&			&			&		&		\\
37838	&	4.3	&	0.94	&	0.33	&		&		&	23.32	&	121	&		&	1.71	&	622	&	30.25	&	-15.39	&		&		-0.43	&			&	d	&		\\
37928	&	3.1	&	1.40	&	0.04	&		&		&	23.70	&	95	&		&	1.86	&	932	&	30.81	&	-17.96	&	0.75	&		-1.05	&			&	d	&		\\
38068	&	4.0	&	1.71	&	0.06	&		&	0.62	&	23.09	&	284	&	1.94	&	2.39	&	710	&	30.60	&	-19.84	&	-0.07	&		-0.86	&		-0.85	&	b,d	&		\\
38150	&	5.3	&	1.61	&	0.39	&	-0.03	&	0.66	&	22.99	&	301	&	1.99	&	2.18	&	769	&	30.63	&	-19.45	&	0.46	&		-0.97	&		-1.20	&	d,f	&		\\
38277	&	1.3	&	0.78	&	0.07	&		&		&	22.68	&	175	&		&	2.11	&	581	&	30.13	&	-15.17	&		&		-0.70	&			&	d	&		\\
38286	&	2.5	&	0.79	&	0.26	&		&		&	23.22	&		&		&		&	538	&	29.97	&	-14.49	&		&			&			&		&		\\
38392	&	3.0	&	1.49	&	0.25	&		&		&	22.81	&	328	&		&	2.23	&	843	&	30.94	&	-19.38	&	-2.52	&		-1.20	&		-0.57	&	b,d	&		\\
38527	&	-1.3	&	1.50	&	0.25	&	0.22	&	0.76	&	23.64	&		&	1.70	&		&	1656	&	31.96	&	-19.64	&		&	$<$	-2.15	&			&	c,d,e	&		\\
38582	&	2.0	&	1.06	&	0.19	&		&		&	23.03	&	153	&		&	1.89	&	946	&	31.08	&	-17.09	&		&		-1.16	&			&	d	&		\\
38800	&	-1.7	&	0.82	&	0.05	&		&		&	22.94	&		&		&		&	1078	&	31.12	&	-16.22	&		&			&			&		&		\\
38802	&	3.0	&	1.25	&	0.09	&		&		&	23.28	&	315	&		&	2.36	&	2460	&	32.83	&	-19.66	&	-0.14	&		-0.61	&		-0.91	&	a,d,e	&		\\
38964	&	2.1	&	1.19	&	0.46	&		&		&	22.37	&	416	&		&	2.30	&	2083	&	32.43	&	-19.73	&	-1.26	&		-1.24	&			&	d	&		\\
39251	&	-0.8	&	1.28	&	0.40	&	0.30	&	0.85	&	23.06	&		&		&		&	2072	&	32.41	&	-19.52	&		&	$<$	-1.60	&			&	c,e	&		\\
39393	&	4.0	&	1.32	&	0.19	&		&	0.80	&	22.51	&	259	&		&	2.16	&	864	&	30.69	&	-18.65	&	-0.34	&		-1.37	&		-1.30	&	b,d	&		\\
39483	&	5.5	&	1.18	&	0.13	&		&	0.63	&	23.47	&	127	&		&	1.86	&	733	&	30.27	&	-16.53	&		&		-0.37	&			&	d	&		\\
39525	&	1.0	&	0.95	&	0.09	&	-0.01	&	0.62	&	22.34	&		&		&		&	3875	&	33.81	&	-20.09	&	-1.88	&			&			&	d	&		\\
39681	&	4.8	&	1.02	&	0.19	&		&		&	24.18	&		&		&		&	925	&	30.85	&	-15.57	&		&			&			&		&		\\
40330	&	5.0	&	1.47	&	0.40	&		&		&	24.48	&	211	&		&	1.97	&	1248	&	31.54	&	-18.20	&	0.32	&		-0.13	&		-1.42	&	d,f	&		\\
40396	&	1.0	&	1.32	&	0.02	&	-0.04	&	0.62	&	22.82	&	134	&		&	2.19	&	1028	&	31.22	&	-18.85	&	-1.20	&		-1.12	&			&	c,d	&		\\
40475	&	1.0	&	1.11	&	0.11	&	-0.28	&	0.40	&	22.68	&	141	&		&	1.91	&	2513	&	32.98	&	-19.70	&	-1.02	&		-0.82	&			&	d	&		\\
\hline 
        \end{tabular}
\end{table}
\clearpage

\begin{table}
            \begin{tabular}{lrrrrrrrrrrrrrrrll}
\hline 
40490	&	1.0	&	1.40	&	0.03	&	0.43	&	0.91	&	23.22	&	350	&	2.28	&	2.64	&	2551	&	32.85	&	-20.46	&		&		-0.79	&			&	d,e	&		\\
40500	&	-3.0	&	1.05	&	0.00	&		&		&	23.00	&		&		&		&	1358	&	31.83	&	-17.96	&		&			&			&	c	&		\\
40715	&	3.2	&	1.16	&	0.05	&		&		&	22.67	&	136	&		&	2.05	&	2295	&	32.62	&	-19.55	&	-1.16	&		-1.01	&			&	d	&		\\
40775	&	4.7	&	1.30	&	0.32	&		&		&	22.51	&	208	&		&	1.98	&	1689	&	31.97	&	-19.72	&	-0.37	&		-0.81	&			&	d	&		\\
41013	&	5.9	&	1.05	&	0.10	&		&		&	22.68	&	134	&		&	1.91	&	1702	&	32.07	&	-18.51	&	0.09	&		-1.06	&			&	d	&		\\
41317	&	5.1	&	1.33	&	0.30	&		&	0.61	&	23.06	&	340	&		&	2.23	&	2443	&	32.76	&	-20.14	&	0.07	&		-0.66	&			&	d	&		\\
41436	&	1.1	&	1.21	&	0.16	&	-0.10	&	0.53	&	22.72	&	332	&		&	2.29	&	3117	&	33.41	&	-20.55	&	-1.18	&		-0.96	&			&	d	&		\\
41652	&	-2.9	&	1.12	&	0.09	&		&		&	22.63	&		&	1.96	&		&	979	&	30.96	&	-17.84	&		&			&			&	c	&		\\
42174	&	1.6	&	1.30	&	0.15	&		&		&	22.90	&	196	&		&	2.07	&	1036	&	30.97	&	-18.39	&	-0.02	&		-1.89	&	$<$	-1.38	&	b,d,a	&		\\
42396	&	3.4	&	1.22	&	0.20	&		&		&	22.43	&	153	&		&	1.89	&	632	&	30.12	&	-17.66	&	0.47	&		-0.96	&			&	a,d	&		\\
42497	&	-1.3	&	0.79	&	0.24	&		&		&	23.82	&		&		&		&	718	&	30.22	&	-14.18	&		&			&			&		&		\\
42605	&	-3.4	&	0.98	&	0.32	&		&		&	23.84	&		&		&		&	1032	&	31.00	&	-15.84	&		&			&			&		&		\\
42833	&	5.2	&	1.60	&	0.18	&		&	0.50	&	22.84	&	384	&		&	2.34	&	804	&	30.59	&	-19.58	&	0.08	&		-0.63	&		-1.07	&	b,d	&	ARP189,VV56	\\
43106	&	6.0	&	1.30	&	0.13	&		&		&	23.90	&	134	&		&	1.89	&	836	&	30.55	&	-16.98	&		&		-0.40	&			&	d	&		\\
43113	&	2.6	&	0.95	&	0.55	&		&		&	23.11	&		&		&		&	4671	&	34.20	&	-19.58	&		&			&			&		&		\\
43121	&	-3.4	&	0.79	&	0.15	&		&		&	23.24	&		&		&		&	830	&	30.75	&	-15.24	&		&			&			&	d	&		\\
43254	&	1.4	&	1.58	&	0.24	&	0.37	&	0.83	&	22.99	&	417	&	2.24	&	2.40	&	1003	&	30.94	&	-19.64	&		&		-0.75	&	$<$	-1.45	&	a,b,d,e	&		\\
43375	&	-0.8	&	1.69	&	0.49	&	0.29	&	0.78	&	23.95	&	333	&		&	2.20	&	1324	&	31.56	&	-19.84	&	-0.62	&		-2.07	&		-1.03	&	b,c,d,e	&		\\
43671	&	-1.6	&	1.74	&	0.27	&	0.36	&	0.82	&	23.52	&	80	&	2.34	&	1.48	&	1395	&	31.55	&	-20.59	&	0.85	&		-2.68	&	$<$	-1.63	&	b,c,d,e,g	&		\\
43931	&	3.0	&	1.22	&	0.15	&		&		&	22.12	&	308	&		&	2.28	&	862	&	30.97	&	-18.72	&	-1.10	&		-0.79	&			&	d	&		\\
44370	&	6.2	&	0.97	&	0.17	&	-0.23	&	0.45	&	23.41	&		&		&	1.96	&	2526	&	32.93	&	-18.15	&		&		-0.53	&	$<$	-0.10	&	a,d	&		\\
44797	&	5.2	&	1.34	&	0.02	&		&	0.50	&	22.35	&	152	&		&	2.28	&	968	&	30.82	&	-19.01	&	-0.62	&		-0.82	&			&	d	&		\\
44961	&	6.4	&	1.50	&	0.77	&		&		&	24.32	&	168	&		&	1.89	&	669	&	30.62	&	-17.44	&		&		0.27	&			&	d	&		\\
45836	&	4.4	&	1.22	&	0.13	&		&		&	23.34	&	263	&		&	2.21	&	2612	&	32.99	&	-19.53	&	-0.80	&		-0.41	&		-0.86	&	a,d	&		\\
45879	&	4.0	&	1.07	&	0.68	&		&		&	23.44	&		&		&		&	3142	&	33.45	&	-19.05	&	-0.84	&			&			&	d	&		\\
45883	&	3.8	&	1.47	&	0.05	&		&		&	24.11	&	243	&		&	2.33	&	3360	&	33.47	&	-20.60	&	-1.55	&		-0.27	&			&	d	&		\\
46934	&	3.3	&	1.31	&	0.07	&		&		&	23.71	&	226	&		&	2.25	&	1225	&	31.59	&	-18.21	&	-2.27	&		-0.30	&			&	d	&		\\
47482	&	3.2	&	0.99	&	0.10	&	0.04	&	0.69	&	22.84	&	298	&		&	2.30	&	4130	&	33.92	&	-19.89	&	-0.25	&		-1.00	&			&	d	&		\\
47577	&	5.9	&	1.06	&	0.17	&		&		&	23.06	&	121	&		&	1.79	&	1020	&	31.10	&	-17.16	&	-0.43	&		-0.62	&			&	d	&		\\
47938	&	-1.6	&	0.90	&	0.49	&		&		&	24.08	&	122	&		&	1.58	&	853	&	30.85	&	-14.99	&		&		0.30	&			&	d	&		\\
47985	&	3.1	&	1.34	&	0.65	&		&		&	24.02	&	257	&		&	2.05	&	1618	&	32.13	&	-18.47	&	-0.80	&		-0.25	&			&	d	&		\\
48425	&	-2.0	&	1.07	&	0.06	&		&		&	23.56	&		&	2.17	&		&	2957	&	33.33	&	-18.99	&		&			&			&		&		\\
48521	&	-1.9	&	1.44	&	0.10	&	0.33	&	0.82	&	23.47	&		&	1.83	&		&	1108	&	31.37	&	-18.90	&		&	$<$	-1.76	&			&	c,e	&		\\
\hline 
        \end{tabular}
\end{table}
\clearpage

\begin{table}
            \begin{tabular}{lrrrrrrrrrrrrrrrll}
\hline
48959	&	5.3	&	1.54	&	0.19	&		&		&	23.83	&	231	&		&	2.09	&	1172	&	31.25	&	-18.94	&	0.41	&		-0.63	&			&	d	&		\\ 
49112	&	5.2	&	1.48	&	0.30	&		&		&	24.27	&	307	&		&	2.19	&	2614	&	33.04	&	-19.89	&	-0.30	&		-0.16	&			&	d	&		\\
49308	&	5.1	&	1.59	&	0.13	&		&		&	24.20	&	240	&		&	2.16	&	1382	&	31.56	&	-19.22	&	0.59	&		-0.43	&			&	d	&		\\
49451	&		&	0.81	&	0.25	&		&		&	21.70	&		&		&		&	1716	&	32.27	&	-18.41	&	-1.68	&			&			&	d	&		\\
49927	&	-1.0	&	0.83	&	0.03	&		&		&	22.78	&		&		&		&	2684	&	33.09	&	-18.27	&		&			&			&		&		\\
49956	&	-0.1	&	1.05	&	0.15	&		&		&	23.39	&	346	&		&	2.31	&	3652	&	33.71	&	-19.37	&		&		-0.07	&			&	d	&		\\
50116	&	3.1	&	1.03	&	0.56	&		&		&	22.18	&		&		&		&	3830	&	33.81	&	-20.46	&		&			&			&		&		\\
50144	&	-2.0	&	1.15	&	0.00	&		&		&	25.31	&		&		&		&	1224	&	31.35	&	-15.70	&		&			&			&		&		\\
50198	&	1.3	&	0.74	&	0.21	&		&		&	23.00	&		&		&		&	4334	&	34.06	&	-18.57	&	-1.97	&			&			&	d	&		\\
50479	&	-1.4	&	1.19	&	0.35	&		&		&	23.91	&		&		&		&	1531	&	31.78	&	-17.73	&		&			&			&		&		\\
50745	&	1.2	&	0.99	&	0.12	&		&		&	21.91	&		&		&		&	4277	&	33.99	&	-20.88	&	-0.05	&			&			&	d	&		\\
50889	&	3.1	&	1.27	&	0.66	&		&		&	23.26	&	451	&		&	2.31	&	4574	&	34.14	&	-20.99	&	-0.86	&		-0.36	&			&	d	&		\\
51091	&	2.8	&	1.07	&	0.40	&		&		&	23.28	&	390	&		&	2.28	&	4280	&	34.03	&	-19.86	&	-0.76	&		-0.20	&			&	d	&		\\
51895	&	-1.9	&	1.10	&	0.38	&		&		&	23.80	&	238	&		&	1.98	&	1454	&	31.71	&	-17.24	&	-2.87	&		0.19	&			&	c,d	&		\\
51951	&	0.1	&	0.60	&	0.03	&		&		&	21.66	&		&		&		&	4281	&	34.03	&	-19.25	&	-1.20	&			&			&	d	&		\\
52273	&	5.4	&	1.52	&	0.52	&		&		&	23.02	&	310	&		&	2.15	&	1753	&	32.09	&	-20.50	&	-0.64	&		-0.69	&			&	d	&		\\
52488	&	3.0	&	1.07	&	0.41	&		&		&	24.16	&		&		&	2.04	&	4605	&	34.14	&	-19.19	&		&		-0.22	&			&	d	&		\\
52607	&	5.9	&	1.18	&	0.30	&		&		&	22.99	&	255	&		&	2.09	&	3284	&	33.53	&	-20.19	&	0.18	&		-1.15	&			&	d	&		\\
52636	&	-3.8	&	0.90	&	0.00	&		&		&	23.93	&		&		&		&	3693	&	33.72	&	-18.30	&		&			&			&		&		\\
52741	&	-1.9	&	0.91	&	0.07	&		&		&	22.90	&		&	2.07	&		&	1693	&	32.18	&	-17.68	&	-0.42	&			&			&	d	&		\\
52887	&	3.8	&	1.17	&	0.13	&		&		&	24.20	&	189	&		&	2.06	&	1781	&	32.18	&	-17.70	&		&		0.00	&			&	d	&		\\
53641	&	2.7	&	0.90	&	0.32	&		&		&	22.10	&	95	&		&	1.56	&	608	&	30.45	&	-16.60	&		&		-0.40	&			&	d	&		\\
54265	&	0.9	&	1.12	&	0.49	&		&		&	23.61	&	453	&		&	2.33	&	3374	&	33.59	&	-19.41	&		&		-0.29	&			&	d	&		\\
54909	&	5.9	&	0.92	&	0.54	&		&		&	23.80	&	181	&		&	1.88	&	1830	&	32.20	&	-16.84	&		&		-0.12	&			&	d	&		\\
55419	&	4.7	&	1.52	&	0.16	&	-0.20	&	0.46	&	24.04	&	215	&		&	2.08	&	655	&	30.63	&	-17.95	&	-0.42	&		-0.07	&			&	d	&		\\
55802	&	6.0	&	0.96	&	0.03	&		&		&	22.27	&		&		&	2.31	&	2864	&	33.27	&	-19.64	&	-0.63	&		-0.85	&			&	d	&		\\
56108	&	5.8	&	1.15	&	0.73	&		&		&	23.16	&	345	&		&	2.20	&	4225	&	33.96	&	-20.34	&	-0.06	&		-0.53	&			&	d	&		\\
56334	&	1.9	&	1.44	&	0.14	&	-0.04	&	0.59	&	23.62	&	178	&		&	1.99	&	1853	&	32.28	&	-19.77	&	0.13	&		-0.46	&			&	d	&		\\
56925	&	3.2	&	1.00	&	0.21	&		&		&	22.30	&		&		&		&	3370	&	33.51	&	-20.13	&		&			&			&		&		\\
57471	&	2.0	&	0.96	&	0.28	&		&		&	23.57	&		&		&		&	3904	&	33.88	&	-18.90	&	-1.15	&			&			&	d	&		\\
58115	&	5.7	&	1.13	&	0.18	&		&		&	22.90	&	201	&		&	2.05	&	2416	&	32.92	&	-19.44	&	-1.28	&		-0.87	&			&	d	&		\\
58183	&	-2.0	&	1.08	&	0.27	&		&		&	23.78	&		&		&		&	2409	&	32.84	&	-18.50	&		&			&			&		&		\\
58336	&	5.8	&	1.22	&	0.45	&		&		&	24.85	&	243	&		&	2.05	&	2988	&	33.28	&	-18.40	&		&		0.31	&			&	d	&		\\
\hline 
        \end{tabular}
\end{table}
\clearpage
\begin{table}
           \begin{tabular}{lrrrrrrrrrrrrrrrll}
\hline
58633	&	4.5	&	1.18	&	0.65	&		&		&	24.34	&	123	&		&	1.68	&	851	&	30.96	&	-16.20	&		&		-0.03	&			&	d	&		\\ 
58827	&	4.9	&	1.47	&	0.38	&	-0.25	&	0.44	&	22.47	&	240	&		&	2.05	&	851	&	30.96	&	-19.59	&	-0.03	&		-0.76	&		-1.47	&	b,d	&		\\
69898	&	5.9	&	1.27	&	0.05	&		&		&	24.73	&	204	&		&	2.27	&	3504	&	33.52	&	-19.27	&		&		-0.22	&			&	d	&		\\
71133	&	1.2	&	1.18	&	0.01	&	0.13	&	0.70	&	22.50	&	207	&		&	2.59	&	1623	&	31.93	&	-19.22	&	-2.03	&		-0.44	&		-0.50	&	b,c,d	&	ARP212,VV280	\\
71360	&	1.0	&	1.03	&	0.60	&		&		&	23.69	&		&		&		&	3033	&	33.21	&	-18.59	&		&			&			&	d	&		\\
71699	&	4.0	&	1.28	&	0.10	&		&		&	23.95	&	228	&		&	2.20	&	4042	&	33.84	&	-20.31	&	-0.06	&		-0.47	&			&	d	&		\\
71796	&	4.0	&	0.87	&	0.19	&		&		&	23.20	&	201	&		&	2.02	&	4168	&	33.91	&	-19.01	&		&		-0.22	&			&	d	&		\\
73163	&	4.2	&	1.14	&	0.04	&	-0.21	&	0.51	&	22.34	&	96	&		&	1.94	&	2404	&	32.75	&	-20.15	&	-1.27	&		-1.15	&			&	d	&		\\
\hline \hline
        \end{tabular}
\vskip 0.5truecm
Notes: The presence of the galaxy in Arp or Vorontsov Velyaminov atlases is reported with the 
original name (ARP and VV). The presence of peculiar kinematics such as counterrotation is indicated
with CR.
REFERENCES:
a: Boselli, A., Gavazzi, G., Lequeux, J., Buat, V., Casoli, F., Dickey, J.\ \& Donas, J.\ 1995, A\&A, 300, L13; 
b: Young, J.\ S.\ Xie Shuding, Tacconi, L. J., et al.\  1995, ApJS, 98, 219;
c: Knapp, G.R., Guhathakurta, P., Kim, D.-W.,Jura, M., 1989, ApJS  70, 329;
d: (LEDA) Paturel, G., Andernach, H., Bottinelli,L., Di Nella, H., Durand, N., Garnier, R., 
    Gouguenheim, L., Lanoix, P., Martinet,M.C., Petit, C., Rousseau, J., Theureau, G., Vauglin, I., 
	1997, A\&AS, 124, 109;
e: Roberts, M., Hogg, D.E., Bregman, J.N., Forman, W.R., Jones, C., 1991, ApJS, 75, 751;
f: Sage, L.\ J.\ 1993, A\&A, 100, 537 and A\&A, 272, 123;  
g: van Driel, W., Ragaigne, D., Boselli, A., Donas, J., \& Gavazzi, G.\ 2000, A\&AS, 144, 463; 
h: Zhu, Ming; Seaquist, E. R.; Davoust, Emmanuel; Frayer, David T.; Bushouse, Howard\ 1999, AJ, 118, 145 
\end{table}
\clearpage
\begin{table}
\setcounter{table}{1}
      \caption[]{Main properties of the Perturbed galaxies. All the propertis were extracted form LEDA except the molecular gas content}
\tabcolsep 5pt
            \begin{tabular}{lrrrrrrrrrrrrrrrll}
PGC	&      t &   log D$_{25}$ 	& log r$_{25}$ & (U-B)$_o$	& (B-V)$_o$ & $\mu_{25}$ 	& W$_{20}$ 	& log $\sigma$ & log v$_m$  &	cz &	m-M$_{cin}$ & M$_B$ & FIR-B & log $\frac{M_{HI}}{L_B}$ & log $\frac{M_{mol}}{L_B}$ & Ref.	&	Notes	\\
\hline
696	&	-0.1	&	1.08	&	0.12	&		&		&	24.31	&		&		&		&	4031	&	33.80	&	-18.79	&		&			&			&		&		\\
2357	&	-1.9	&	1.37	&	0.42	&	0.36	&	0.84	&	24.56	&		&		&		&	4231	&	33.89	&	-20.02	&		&			&			&		&		\\
2365	&	-1.2	&	0.97	&	0.10	&	0.05	&	0.59	&	23.52	&		&	2.30	&		&	4154	&	33.85	&	-19.03	&		&			&			&		&		\\
4777	&	3.1	&	1.45	&	0.26	&	0.02	&	0.65	&	23.12	&	394.08	&		&	2.32	&	2374	&	32.63	&	-20.63	&	-0.89	&		-0.78	&	$<$	-0.96	&	b,d	&		\\
4801	&	-2.0	&	1.79	&	0.09	&	0.35	&	0.80	&	25.03	&	367.85	&	2.23	&	2.41	&	2365	&	32.62	&	-20.44	&		&		-1.32	&			&	c,d,e	&	ARP227	\\
7525	&	3.0	&	1.85	&	0.25	&	0.15	&	0.65	&	23.80	&	472.06	&	2.15	&	2.47	&	2456	&	32.76	&	-22.23	&	0.38	&		-0.66	&		-0.60	&	b,d	&	ARP78	\\
7533	&	6.5	&	1.15	&	0.33	&		&		&	24.87	&	190.50	&		&	1.96	&	4613	&	34.10	&	-18.92	&		&		-0.34	&			&	d	&		\\
7846	&	-1.9	&	1.18	&	0.24	&	0.41	&	0.85	&	23.72	&		&		&		&	3452	&	33.47	&	-19.66	&		&			&			&		&	ARP290,VV309	\\
7856	&	3.1	&	1.40	&	0.42	&		&		&	23.97	&	381.50	&	2.10	&	2.26	&	3646	&	33.59	&	-20.53	&		&		-0.82	&			&	d	&	ARP290,VV309	\\
8360	&	6.0	&	1.34	&	0.20	&		&		&	24.73	&	231.57	&		&	2.09	&	3256	&	33.31	&	-19.17	&		&		0.09	&			&	d	&		\\
26232	&	1.1	&	1.43	&	0.46	&	-0.09	&	0.60	&	23.45	&	316.11	&		&	2.12	&	1739	&	32.17	&	-19.62	&	-2.66	&		-0.82	&		-0.42	&	b,d,e	&	ARP283,VV50	\\
26498	&	5.4	&	1.65	&	1.04	&		&	0.23	&	23.41	&	370.13	&		&	2.22	&	1580	&	32.07	&	-20.70	&	0.21	&		-0.18	&			&	d	&		\\
26571	&	1.0	&	1.17	&	0.03	&		&		&	23.67	&	230.35	&	2.28	&	2.37	&	1782	&	32.21	&	-18.19	&		&		-0.47	&			&	d	&		\\
26580	&	-0.9	&	1.27	&	0.25	&		&		&	24.45	&		&		&		&	1765	&	32.19	&	-17.88	&		&			&			&		&		\\
27159	&	-2.1	&	1.57	&	0.17	&	0.43	&	0.89	&	24.27	&	518.66	&	2.37	&	2.40	&	3179	&	33.30	&	-20.79	&		&		-1.17	&			&	c,d,e	&	ARP232	\\
27939	&	1.6	&	1.16	&	0.06	&		&		&	23.27	&		&		&	2.39	&	4430	&	34.13	&	-20.87	&	0.09	&		-1.29	&			&	d	&		\\
28197	&	5.8	&	1.21	&	0.83	&		&		&	23.56	&		&		&	2.04	&	3025	&	33.35	&	-19.68	&		&		-0.25	&			&	d	&		\\
29814	&	0.2	&	1.64	&	0.29	&	0.32	&	0.84	&	23.50	&	139.08	&	2.11	&	1.82	&	1331	&	31.39	&	-19.92	&	-0.34	&		-1.59	&		-1.29	&	c,d,e	&		\\
29855	&	1.2	&	1.67	&	0.27	&	0.19	&	0.76	&	22.90	&	508.46	&	2.23	&	2.25	&	1233	&	31.22	&	-20.50	&	0.09	&		-0.63	&		-0.61	&	b,d	&		\\
30068	&	5.0	&	1.38	&	0.41	&	-0.28	&	0.36	&	23.85	&	262.37	&		&	2.07	&	1578	&	31.88	&	-18.72	&	-0.16	&		-0.52	&			&	d	&	ARP316,VV307	\\
30083	&	0.9	&	1.63	&	0.41	&	0.38	&	0.86	&	23.58	&	527.87	&	2.24	&	2.39	&	1312	&	31.51	&	-19.87	&	-0.10	&		-1.48	&			&	c,d,e	&	ARP316,VV307	\\
30445	&	1.4	&	1.70	&	0.23	&	0.21	&	0.75	&	23.32	&	418.71	&	2.13	&	2.35	&	1154	&	31.23	&	-20.19	&	-0.25	&		-1.09	&			&	d	&	ARP94,VV209	\\
30714	&	3.4	&	1.50	&	0.99	&		&		&	25.13	&	198.50	&		&	1.90	&	1324	&	31.58	&	-17.59	&		&		0.07	&			&	d	&		\\
32292	&	-2.6	&	1.73	&	0.30	&	0.40	&	0.88	&	23.15	&	84.75	&	2.15	&	1.38	&	890	&	30.65	&	-19.97	&		&		-2.15	&			&	c,d,e,g	&		\\
32306	&	5.3	&	1.45	&	0.35	&	-0.23	&	0.36	&	22.70	&	275.17	&		&	2.15	&	1291	&	31.42	&	-19.78	&		&		-0.79	&			&	d	&		\\
32533	&	-2.1	&	1.50	&	0.14	&	0.52	&	0.92	&	23.32	&	324.42	&	2.39	&	2.27	&	1524	&	31.86	&	-19.92	&		&		-1.86	&			&	c,d,e,g	&	ARP162	\\
32584	&	3.1	&	1.44	&	0.55	&		&		&	23.14	&	399.35	&		&	2.24	&	1503	&	31.86	&	-19.66	&	-1.48	&		-0.69	&			&	d	&		\\
32605	&	5.2	&	1.52	&	0.03	&		&		&	24.41	&	271.48	&		&	2.54	&	2719	&	32.98	&	-20.07	&		&		-0.21	&			&	d	&		\\
32767	&	3.2	&	1.42	&	0.27	&		&		&	23.95	&	215.36	&		&	2.02	&	1107	&	31.14	&	-18.11	&		&		-0.28	&			&	d	&		\\
34029	&	6.0	&	1.09	&	0.06	&		&		&	24.58	&	124.23	&		&	1.97	&	3052	&	33.36	&	-18.01	&		&		0.24	&			&	d	&		\\
\hline 
        \end{tabular}
\end{table}
\clearpage
\begin{table}
           \begin{tabular}{lrrrrrrrrrrrrrrrll}
\hline 
34415	&	-4.6	&	0.97	&	0.20	&	0.44	&	0.83	&	21.89	&		&	1.99	&		&	661	&	30.15	&	-17.07	&		&	$<$	-2.03	&			&	e	&		\\
34561	&	5.3	&	1.62	&	0.18	&		&		&	24.42	&	300.53	&		&	2.24	&	2331	&	32.80	&	-20.23	&	0.36	&		-0.25	&			&	d	&		\\
35616	&	1.1	&	1.88	&	0.27	&	0.23	&	0.74	&	24.39	&	474.00	&	2.25	&	2.40	&	992	&	31.21	&	-19.97	&	1.88	&		-0.33	&	$<$	-1.22	&	b,c,d	&	ARP214	\\
35999	&	3.4	&	1.44	&	0.50	&		&		&	22.73	&	255.24	&		&	2.06	&	730	&	30.66	&	-18.89	&		&		-0.40	&			&	d	&	ARP280	\\
36060	&	6.4	&	1.11	&	0.33	&	-0.18	&	0.39	&	23.72	&	222.94	&		&	2.03	&	3358	&	33.48	&	-19.09	&		&		-0.27	&			&	d	&		\\
36158	&	1.1	&	1.30	&	0.28	&		&		&	23.66	&	450.89	&	2.15	&	2.35	&	2718	&	33.07	&	-19.74	&		&		-0.36	&			&	d	&	ARP294,VV228	\\

36160	&	2.2	&	1.23	&	0.57	&		&		&	22.76	&	541.72	&		&	2.42	&	2689	&	33.05	&	-20.21	&		&		-0.57	&			&	d	&	ARP294,VV228	\\
36193	&	3.1	&	0.85	&	0.21	&		&		&	21.94	&	438.28	&		&	2.40	&	3312	&	33.43	&	-19.62	&		&		-0.71	&			&	d	&	ARP83,VV350	\\
36197	&	3.1	&	1.24	&	0.52	&		&		&	22.89	&	430.32	&		&	2.31	&	3306	&	33.43	&	-20.56	&	-1.59	&		-0.97	&			&	d	&	ARP83,VV350	\\
36200	&	-1.8	&	1.46	&	0.22	&	0.35	&	0.84	&	23.99	&	329.75	&	2.25	&	2.21	&	3319	&	33.44	&	-20.65	&		&		-0.98	&			&	c,d	&		\\
36871	&	4.7	&	0.78	&	0.51	&		&		&	22.63	&		&		&		&	4394	&	34.03	&	-19.12	&		&			&			&		&		\\
36897	&	0.2	&	1.18	&	0.16	&		&	0.49	&	23.58	&	277.90	&		&	2.18	&	955	&	31.13	&	-17.25	&		&		0.42	&			&	d	&		\\
36907	&	1.1	&	1.12	&	0.17	&	0.26	&	0.86	&	23.14	&		&		&		&	3301	&	33.53	&	-19.89	&		&			&			&	3	&		\\
37466	&	4.0	&	1.57	&	0.57	&	-0.14	&	0.54	&	23.61	&	267.13	&		&	2.07	&	845	&	30.96	&	-18.90	&	0.65	&		-0.80	&			&	d	&		\\
37618	&	-2.7	&	1.13	&	0.19	&	0.36	&	0.85	&	22.85	&		&	2.25	&		&	695	&	30.64	&	-17.26	&		&			&			&	c	&		\\
37619	&	3.1	&	1.21	&	0.54	&		&		&	23.22	&	374.18	&		&	2.24	&	4826	&	34.25	&	-20.86	&		&		-0.74	&		0.07	&	b,d	&		\\
37629	&	3.0	&	1.16	&	0.28	&	-0.15	&	0.47	&	23.35	&	280.67	&		&	2.13	&	4768	&	34.23	&	-20.49	&	-0.82	&		-0.39	&		-1.59	&	a,d	&		\\
37642	&	-2.1	&	1.46	&	0.08	&	0.50	&	0.92	&	22.59	&	633.00	&	2.48	&	2.67	&	1042	&	31.32	&	-19.83	&		&		-1.41	&			&	c,d,e	&		\\
37692	&	5.9	&	1.35	&	0.09	&	-0.07	&	0.52	&	23.29	&	215.83	&		&	2.15	&	767	&	30.72	&	-18.00	&		&		-0.38	&			&	d	&		\\
37719	&	2.2	&	1.41	&	0.37	&	0.26	&	0.91	&	23.21	&	358.40	&	2.12	&	2.24	&	699	&	30.56	&	-18.15	&		&		-0.54	&			&	d	&		\\
38287	&	1.3	&	1.02	&	0.50	&		&		&	22.73	&	411.05	&	2.24	&	2.28	&	4275	&	33.98	&	-20.18	&		&		-0.48	&			&	d	&		\\
38503	&	-2.0	&	1.29	&	0.41	&		&		&	24.28	&	283.50	&	1.98	&	2.09	&	933	&	31.06	&	-16.97	&		&		-0.01	&			&	d	&		\\
38885	&	1.9	&	1.08	&	0.22	&	0.14	&	0.75	&	23.28	&	270.24	&		&	2.14	&	1862	&	32.22	&	-18.23	&		&		-1.24	&			&	d	&		\\
38892	&	-1.9	&	1.12	&	0.27	&	0.42	&	0.87	&	22.62	&	476.10	&	2.34	&	2.36	&	3814	&	33.77	&	-20.60	&		&		-0.86	&			&	c,d	&		\\
38906	&	-0.2	&	0.84	&	0.35	&		&		&	22.31	&	423.57	&		&	2.19	&	3945	&	33.84	&	-19.54	&		&		-0.04	&			&	d,g	&		\\
38912	&	3.2	&	1.14	&	0.63	&		&		&	22.74	&	261.72	&		&	2.03	&	3941	&	33.84	&	-20.53	&	-2.06	&		-0.60	&		-0.58	&	a,d	&		\\
39568	&	3.0	&	1.61	&	0.78	&		&		&	23.52	&	545.50	&		&	2.42	&	2529	&	33.01	&	-21.28	&	1.73	&		-1.23	&			&	d	&		\\
39687	&	-1.0	&	0.95	&	0.11	&	0.44	&	0.82	&	22.22	&		&		&		&	2632	&	32.92	&	-19.30	&		&			&			&	c	&		\\
39708	&	-1.9	&	0.99	&	0.34	&		&		&	23.71	&	393.90	&		&	2.26	&	2291	&	32.63	&	-17.66	&		&		-0.12	&			&	d	&		\\
39712	&	-0.2	&	1.17	&	0.37	&	0.41	&	0.94	&	23.09	&	292.41	&		&	2.07	&	2374	&	32.70	&	-19.23	&		&		-0.51	&			&	c,d	&		\\
39719	&	-1.0	&	1.12	&	0.14	&		&		&	23.16	&		&		&		&	2208	&	32.55	&	-18.83	&		&			&			&	c	&		\\
39759	&	0.2	&	0.91	&	0.27	&		&	0.92	&	23.11	&	278.75	&		&	2.13	&	2504	&	32.81	&	-18.04	&		&		-0.03	&			&	d	&		\\
\hline 
        \end{tabular}
\end{table}
\clearpage
\begin{table}
           \begin{tabular}{lrrrrrrrrrrrrrrrll}
\hline 
39943	&	-2.0	&	1.03	&	0.27	&		&	0.73	&	23.10	&		&	2.29	&		&	4229	&	33.94	&	-19.85	&		&			&			&	c	&		\\
39950	&	5.2	&	1.47	&	0.26	&		&	0.61	&	22.64	&	264.75	&		&	2.11	&	1140	&	31.24	&	-19.78	&	-0.42	&		-1.11	&		-0.62	&	b,d	&		\\
39974	&	5.4	&	1.69	&	0.82	&		&	0.72	&	23.25	&	376.58	&		&	2.24	&	1118	&	31.20	&	-20.15	&		&		-0.60	&		-0.81	&	b,d	&		\\
39981	&	2.4	&	1.43	&	0.28	&		&		&	23.57	&		&		&		&	1484	&	32.01	&	-19.45	&		&			&			&	d	&		\\
40030	&	1.0	&	1.30	&	0.26	&		&	0.59	&	23.30	&		&		&		&	1893	&	32.26	&	-19.34	&		&			&			&		&		\\
40032	&	-2.0	&	1.16	&	0.16	&		&		&	23.53	&		&		&		&	1650	&	31.98	&	-18.16	&		&			&			&		&		\\
40245	&	-1.2	&	1.51	&	0.11	&	0.46	&	0.88	&	23.36	&		&	2.07	&		&	917	&	30.83	&	-18.85	&		&	$<$	-2.23	&			&	c,e	&		\\
40295	&	-1.8	&	1.46	&	0.28	&	0.42	&	0.88	&	22.91	&		&	2.29	&		&	1226	&	31.40	&	-19.63	&		&	$<$	-2.26	&			&	c,e	&		\\
40309	&	-1.8	&	1.06	&	0.06	&		&		&	23.76	&		&		&		&	4496	&	34.16	&	-19.57	&		&			&			&		&		\\
40515	&	-1.3	&	1.85	&	0.11	&	0.38	&	0.84	&	23.02	&		&	2.26	&		&	742	&	30.44	&	-20.55	&		&	$<$	-3.13	&			&	c,e	&		\\
40581	&	2.8	&	1.74	&	0.59	&	0.15	&	0.58	&	23.41	&	386.07	&	1.93	&	2.33	&	2519	&	32.86	&	-21.92	&	-0.34	&		-1.53	&		-1.28	&	a,b,d	&		\\
40614	&	2.9	&	1.54	&	0.02	&	0.18	&	0.81	&	23.12	&	177.64	&	2.15	&	2.32	&	921	&	30.85	&	-19.30	&	1.09	&		-1.35	&		-0.95	&	b,d	&		\\
40836	&	-1.0	&	1.59	&	0.11	&	0.26	&	0.61	&	25.04	&		&	2.20	&		&	2662	&	33.11	&	-19.87	&	-1.06	&			&			&	c,d	&		\\
40903	&	-1.8	&	1.17	&	0.27	&	0.41	&	0.86	&	23.48	&		&		&		&	1211	&	31.34	&	-17.55	&		&			&			&		&		\\
41302	&	-1.8	&	1.18	&	0.06	&	0.47	&	0.92	&	23.15	&		&	1.91	&		&	850	&	30.65	&	-17.29	&		&			&			&	c	&		\\
41363	&	0.1	&	1.56	&	0.50	&	0.30	&	0.76	&	24.43	&		&		&		&	994	&	30.91	&	-18.05	&		&			&			&	c	&		\\
42064	&	4.0	&	1.44	&	0.13	&	0.26	&	0.68	&	22.82	&	320.95	&		&	2.25	&	2265	&	32.63	&	-20.88	&		&		-0.78	&		-0.62	&	b,d	&	VV219	\\
42069	&	4.1	&	1.64	&	0.34	&		&	0.76	&	22.98	&		&		&	2.27	&	2255	&	32.62	&	-21.66	&	-1.07	&		-1.39	&		-0.61	&	b,d	&	VV219	\\
42620	&	-4.3	&	1.32	&	0.12	&	0.09	&	0.61	&	23.52	&		&		&		&	765	&	30.65	&	-17.69	&		&			&			&	1	&	ARP281	\\
42710	&	8.5	&	0.71	&	0.17	&		&		&	23.30	&	160.75	&		&	1.90	&	1100	&	31.17	&	-15.22	&		&		0.14	&			&	d	&		\\
42728	&	-2.7	&	1.39	&	0.16	&	0.40	&	0.89	&	22.80	&		&	2.11	&		&	1127	&	31.20	&	-19.22	&		&	$<$	-2.12	&			&	e	&		\\
42816	&	5.2	&	1.45	&	0.08	&	0.25	&	0.60	&	22.76	&	212.45	&		&	2.09	&	1415	&	31.66	&	-19.99	&	-0.57	&		-1.22	&		-0.57	&	b,d	&	ARP116,VV206	\\
47777	&	4.9	&	1.05	&	0.62	&		&		&	23.29	&		&		&		&	4600	&	34.15	&	-19.85	&	-1.66	&			&			&	d	&		\\
47867	&	3.3	&	1.05	&	0.10	&		&	0.37	&	24.24	&	226.84	&		&	2.17	&	4962	&	34.34	&	-19.16	&		&		0.31	&			&	d	&	ARP183	\\
48018	&		&	0.84	&	0.34	&		&		&	24.19	&		&		&		&	3886	&	33.75	&	-17.61	&		&			&			&		&		\\
48811	&	-1.1	&	0.98	&	0.22	&	0.35	&	0.56	&	23.88	&		&	2.43	&		&	7333	&	35.17	&	-20.05	&		&			&			&		&		\\
48815	&	4.9	&	1.71	&	0.63	&	-0.06	&	0.49	&	23.60	&	415.44	&	2.21	&	2.29	&	2403	&	32.88	&	-21.54	&	0.58	&		-0.50	&			&	d	&	CR	\\
48860	&	-2.0	&	1.56	&	0.74	&		&	0.82	&	23.74	&		&	2.42	&	2.18	&	2021	&	32.58	&	-20.38	&		&		-1.58	&			&	c,d,e	&		\\
49347	&	3.6	&	1.43	&	0.14	&	0.24	&	0.76	&	23.00	&	309.40	&		&	2.28	&	2315	&	32.80	&	-20.74	&	-0.21	&		-0.53	&			&	d	&		\\
49354	&	-2.1	&	1.32	&	0.18	&	0.49	&	0.91	&	22.83	&	288.90	&		&	2.16	&	2305	&	32.79	&	-20.38	&		&		-0.56	&			&	c,d	&	CR	\\
49356	&	-2.1	&	1.30	&	0.20	&	0.58	&	0.92	&	22.38	&	295.90	&	2.45	&	2.18	&	2305	&	32.79	&	-20.73	&		&		-0.74	&			&	c,d,e	&		\\
49508	&	-1.8	&	1.35	&	0.36	&		&		&	24.56	&	158.50	&		&	1.80	&	1762	&	32.32	&	-18.27	&		&		-0.04	&			&	d	&		\\
\hline 
        \end{tabular}
\end{table}
\clearpage
\begin{table}
           \begin{tabular}{lrrrrrrrrrrrrrrrll}
\hline 
49548	&	-0.4	&	1.60	&	0.56	&	0.34	&	0.78	&	24.44	&		&	2.22	&		&	1842	&	32.40	&	-19.68	&	1.15	&			&			&	c,d	&		\\
49739	&	3.1	&	1.28	&	0.26	&	0.06	&	0.54	&	23.46	&	595.50	&		&	2.50	&	3472	&	33.61	&	-20.30	&	-2.29	&		-0.22	&		-0.39	&	d,a	&	ARP84,VV48	\\
49747	&	3.2	&	1.41	&	0.28	&	0.04	&	0.53	&	22.42	&	625.93	&		&	2.51	&	3486	&	33.62	&	-21.98	&		&		-0.70	&		-0.62	&	d,a	&	ARP84,VV48	\\
49820	&	3.0	&	1.47	&	0.57	&		&		&	23.84	&	512.13	&		&	2.39	&	2743	&	33.13	&	-20.35	&	-1.18	&		-0.23	&			&	d	&	VV310	\\
49824	&	0.1	&	0.63	&	0.28	&		&		&	22.11	&		&		&		&	2748	&	33.14	&	-17.93	&		&			&			&		&	VV310	\\
49893	&	4.9	&	1.13	&	0.28	&		&		&	23.85	&	262.85	&		&	2.11	&	3721	&	33.76	&	-19.34	&	-1.15	&		-0.09	&		-0.62	&	d,a	&	VV256	\\
50331	&	-4.0	&	1.26	&	0.13	&		&		&	23.81	&		&	2.17	&		&	1885	&	32.43	&	-18.90	&		&			&			&	c	&		\\
50776	&	-2.0	&	1.28	&	0.23	&		&		&	24.02	&		&	2.14	&		&	4991	&	34.33	&	-20.64	&		&			&			&		&		\\
51223	&	3.0	&	1.56	&	0.61	&	0.04	&	0.68	&	23.76	&	254.06	&		&	2.02	&	1728	&	32.06	&	-19.84	&	0.03	&		-1.06	&			&	d	&	ARP286	\\
51233	&	1.5	&	1.82	&	0.46	&	0.34	&	0.78	&	23.69	&	504.51	&		&	2.43	&	1506	&	31.78	&	-20.99	&	1.65	&		-1.36	&			&	d,e	&	ARP286	\\
51241	&	5.8	&	1.15	&	0.09	&		&		&	24.75	&	105.53	&		&	1.83	&	1773	&	32.12	&	-16.98	&		&		0.14	&			&	d	&	ARP286	\\
51270	&	-2.7	&	1.16	&	0.17	&	0.28	&	0.78	&	23.04	&	613.60	&		&	2.38	&	1656	&	31.97	&	-18.63	&		&		-1.55	&			&	c,d,e	&		\\
51668	&	-1.9	&	1.01	&	0.19	&		&		&	23.57	&	183.00	&	2.24	&	1.93	&	4370	&	34.07	&	-19.42	&		&		-1.17	&			&	d	&		\\
51681	&	-2.0	&	1.24	&	0.02	&		&	0.90	&	23.13	&		&	2.43	&		&	4518	&	34.14	&	-21.14	&		&			&			&	c	&		\\
51785	&	-0.4	&	1.15	&	0.17	&		&		&	23.86	&		&		&	2.31	&	1676	&	32.00	&	-17.74	&		&		-1.02	&			&	d	&		\\
51965	&	3.2	&	1.42	&	0.48	&		&		&	23.69	&		&		&		&	4068	&	33.94	&	-21.08	&	0.52	&			&			&	d	&		\\
52686	&	3.0	&	1.22	&	0.06	&		&		&	23.57	&	245.90	&	2.24	&	2.31	&	4400	&	34.10	&	-20.42	&		&		-0.58	&			&	d	&	ARP297	\\
53176	&	4.5	&	1.08	&	0.26	&		&		&	23.24	&	219.05	&	2.03	&	2.03	&	1576	&	31.88	&	-17.96	&		&		-0.36	&			&	d	&		\\
53178	&	3.0	&	1.30	&	0.09	&		&		&	23.36	&	223.72	&		&	2.20	&	1577	&	31.88	&	-18.97	&		&		-0.72	&			&	d	&		\\
53217	&	5.2	&	1.46	&	0.21	&		&		&	24.60	&	263.17	&		&	2.15	&	2334	&	32.85	&	-19.43	&	0.38	&		-0.17	&			&	d	&		\\
53657	&	4.0	&	1.24	&	0.68	&		&		&	23.34	&	312.85	&		&	2.13	&	2544	&	33.01	&	-19.58	&	-1.09	&		0.16	&			&	d	&		\\
53995	&	2.5	&	1.08	&	0.29	&	-0.13	&	0.64	&	22.52	&		&	2.22	&	2.38	&	4759	&	34.24	&	-20.98	&		&			&			&		&		\\
54001	&	4.0	&	1.44	&	0.55	&	0.07	&	0.65	&	23.27	&	479.70	&		&	2.37	&	4756	&	34.24	&	-21.98	&	-0.29	&		-0.75	&			&	d	&		\\
55647	&	4.3	&	1.43	&	0.89	&	0.23	&	0.76	&	23.30	&		&		&	2.42	&	2527	&	33.02	&	-20.54	&	0.75	&		-0.45	&			&	d	&		\\
55725	&	3.1	&	1.71	&	0.30	&	0.08	&	0.68	&	23.73	&	534.80	&		&	2.49	&	2519	&	33.01	&	-21.61	&	0.99	&		-0.79	&			&	d	&		\\
57579	&	-3.3	&	0.92	&	0.06	&		&		&	23.26	&		&		&		&	9444	&	35.72	&	-21.04	&		&			&			&		&		\\
57627	&	5.1	&	1.34	&	0.80	&		&		&	23.67	&	226.50	&		&	1.99	&	2306	&	32.74	&	-19.59	&		&		-0.63	&			&	d	&		\\
69630	&	0.3	&	1.22	&	0.18	&		&		&	23.78	&	394.22	&		&	2.35	&	4851	&	34.21	&	-20.59	&		&		-0.42	&			&	d	&		\\
70348	&	1.1	&	1.16	&	0.12	&	-0.46	&	0.48	&	22.38	&	386.28	&		&	2.41	&	4916	&	34.25	&	-21.77	&	-2.79	&		-1.35	&		-0.64	&	b,d	&	ARP298	\\
70786	&	3.6	&	1.31	&	0.59	&	-0.19	&	0.46	&	23.14	&	363.39	&	2.07	&	2.16	&	2676	&	32.93	&	-20.29	&		&		-0.28	&			&	d	&		\\
70795	&	4.7	&	1.52	&	0.46	&	-0.06	&	0.53	&	22.90	&	469.93	&	1.83	&	2.33	&	2683	&	32.93	&	-21.60	&	-1.44	&		-0.57	&		-0.67	&	b,d	&		\\
71034	&	6.8	&	1.16	&	0.50	&	-0.24	&	0.37	&	23.97	&	227.62	&		&	1.97	&	4198	&	33.91	&	-19.84	&		&		-0.50	&			&	d	&		\\
\hline 
        \end{tabular}
\end{table}
\clearpage

\begin{table}
            \begin{tabular}{lrrrrrrrrrrrrrrrll}
\hline 
71113	&	-1.9	&	0.97	&	0.14	&	0.26	&	0.86	&	23.59	&		&	2.06	&		&	4094	&	33.86	&	-19.27	&		&			&			&	c	&		\\
71868	&	3.1	&	1.28	&	0.17	&	-0.52	&	0.42	&	22.97	&	235.32	&		&	2.07	&	2799	&	33.01	&	-20.42	&	-1.86	&		-0.50	&		-0.87	&	b,d	&	ARP284,VV51	\\
72128	&	1.0	&	1.11	&	0.10	&	-0.03	&	0.66	&	23.86	&	206.58	&		&	2.11	&	2892	&	33.09	&	-18.77	&	-1.22	&		0.11	&			&	d	&		\\
72638	&	1.1	&	1.38	&	0.31	&	0.24	&	0.67	&	23.35	&	609.69	&		&	2.49	&	4300	&	33.98	&	-21.58	&	-2.49	&		-0.79	&			&	d	&		\\
\hline \hline
        \end{tabular}
\vskip 0.5truecm
Notes: The presence of the galaxy in Arp or Vorontsov Velyaminov atlases is reported with the 
original name (ARP and VV). The presence of peculiar kinematics such as counterrotation is indicated
with CR.
REFERENCES:
a: Boselli, A., Gavazzi, G., Lequeux, J., Buat, V., Casoli, F., Dickey, J.\ \& Donas, J.\ 1995, A\&A, 300, L13; 
b: Young, J.\ S.\ Xie Shuding, Tacconi, L. J., et al.\  1995, ApJS, 98, 219;
c: Knapp, G.R., Guhathakurta, P., Kim, D.-W.,Jura, M., 1989, ApJS  70, 329;
d: (LEDA) Paturel, G., Andernach, H., Bottinelli,L., Di Nella, H., Durand, N., Garnier, R., 
    Gouguenheim, L., Lanoix, P., Martinet,M.C., Petit, C., Rousseau, J., Theureau, G., Vauglin, I., 
	1997, A\&AS, 124, 109;
e: Roberts, M., Hogg, D.E., Bregman, J.N., Forman, W.R., Jones, C., 1991, ApJS, 75, 751;
f: Sage, L.\ J.\ 1993, A\&A, 100, 537 and A\&A, 272, 123;  
g: van Driel, W., Ragaigne, D., Boselli, A., Donas, J., \& Gavazzi, G.\ 2000, A\&AS, 144, 463; 
h: Zhu, Ming; Seaquist, E. R.; Davoust, Emmanuel; Frayer, David T.; Bushouse, Howard\ 1999, AJ, 118, 145 
\end{table}
\end{landscape}

\end{document}